\documentclass[11pt,a4paper]{article}
\usepackage{amssymb,amsmath, amsfonts}
\usepackage{graphicx,graphics}
\usepackage{mathtools}
\usepackage{bbold}
\usepackage{comment}
\usepackage[english]{babel}
\usepackage[utf8]{inputenc} 
\usepackage{epsfig,url}
\usepackage{bbm,theorem}
\usepackage{a4wide}
\usepackage{color} 
\usepackage{enumerate}
\usepackage{calrsfs}
\usepackage[bookmarks=true]{hyperref}
\usepackage{bookmark}
\usepackage{amsmath}
\usepackage{amsfonts}
\usepackage{amssymb}
\usepackage{graphicx}

\providecommand{\U}[1]{\protect\rule{.1in}{.1in}}
\DeclareMathAlphabet{\pazocal}{OMS}{zplm}{m}{n}
\newtheorem{theorem}{Theorem}

\newtheorem{conjecture}[theorem]{Conjecture}

\newtheorem{remark}[theorem]{Remark}

\numberwithin{equation}{section}
\numberwithin{theorem}{section}

\newcommand{\R}{{\mathbb R}}

\newcommand{\N}{{\mathbb N}}

\newcommand{\EE}{\mathbb{E}}
\newcommand{\ep}{\varepsilon}

\newcommand{\ud}{\;\mathrm{d}}

\newcommand{\eps}{{\varepsilon}}

\newcommand{\beq}{\begin{equation}}
\newcommand{\eeq}{\end{equation}}
\newcommand{\beqs}{\begin{eqnarray}}
\newcommand{\eeqs}{\end{eqnarray}}

\newcounter{jlisti}

\newcommand{\func}[1]{\operatorname{#1}}

\def\be{\begin{equation}}
\def\ee{\end{equation}}
\def\bea{\begin{eqnarray}}
\def\eea{\end{eqnarray}}

\begin{document}

\title{Interacting particle systems
with long-range interactions: scaling limits and kinetic equations } 

\author{Alessia Nota \thanks{\emailalessia} , Juan J. L. Vel\'azquez \thanks{\emailjuan}, Raphael Winter \thanks{\emailraphael} \\[1em]
$\,^*$\UAaddress\\[0.5em] $\, ^\dag$\UBaddress \\[0.5em]
 $\,^\ddag$\ULaddress}

\date{\today}

\newcommand{\email}[1]{E-mail: \tt #1}
\newcommand{\emailalessia}{\email{nota@iam.uni-bonn.de} (corresponding author)}
\newcommand{\emailjuan}{\email{velazquez@iam.uni-bonn.de}}
\newcommand{\emailraphael}{\email{raphael.winter@ens-lyon.fr}}

\newcommand{\UAaddress}{\em Dipartimento di Ingegneria e Scienze dell'Informazione e Matematica,\\ \em Universit\`a degli studi dell'Aquila, L'Aquila, 67100 Italy}
\newcommand{\UBaddress}{\em University of Bonn, Institute for Applied Mathematics\\
\em Endenicher Allee 60, D-53115 Bonn, Germany}
\newcommand{\ULaddress}{\em Universit\'e de Lyon, \\ \em 43 Boulevard du 11 Novembre 1918, 69100 Villeurbanne, France}

\date{\today }
\maketitle

\begin{abstract}
The goal of this paper is to describe the various kinetic equations which arise from scaling limits of interacting particle systems. We provide a formalism which allows us to determine the kinetic equation for a given interaction potential and scaling limit. Our focus in this paper is on particle systems with long-range interactions. The derivation here is formal, but it provides an interpretation of particle systems as the motion of a particle in a random force field with a friction term which is due to the interaction with the surrounding
particles. Some of the  technical details of this method are discussed in the companion paper \cite{NVW}.
\end{abstract}

\tableofcontents

\bigskip\ \noindent

\vspace{0.5cm}

\bigskip

\section{Introduction} 
It is well known that a large class of many particle systems which evolve by
means of Newton equations can be described, under suitable assumptions on
the potentials describing the particle interactions, by means of a function $%
f\left( x,v,t\right) $ which yields the density of particles in the phase
space $\left( x,v\right) \in\mathbb{R}^{3}\times\mathbb{R}^{3}.$ The
evolution of the function $f\left( x,v,t\right) ,$ which is usually termed
as one-particle distribution function, in these cases is given by a kinetic
equation. Some examples of kinetic equations are the Boltzmann equation, the
Landau equation and the Balescu-Lenard equation.

The specific kinetic equation which describes a given scaling limit of
interacting particles systems depends on the properties of the interaction potentials
as well as on the scalings assumed for the magnitudes describing the interaction,
like the strength and range of the potential as well as other properties of
the potential which will be described in detail in this paper and its companion paper \cite{NVW}.

One of the main goals of kinetic theory is to describe, the mesoscopic behavior of a system of particles
whose evolution is given by the Newton equations:%
\begin{equation}
\frac{dX_{j}}{d\tau}=V_{j}\ \ \ ,\ \ \frac{dV_{j}}{d\tau}=-\sum_{k}\nabla
\Phi_{\varepsilon}\left( X_{j}-X_{k}\right) \ \ ,\ \ j\in S  \label{S1E1}
\end{equation}
where $S$ is a (countable) set of indexes. The potentials $\Phi_{\varepsilon
}\left( X\right) =\Phi\left( X;\varepsilon\right) $ depend on a parameter $%
\varepsilon$ which eventually will be sent to zero. The problem is to
characterize the families of potentials $\Phi_{\varepsilon}$ for which it is
possible to describe the evolution of the system by means of an equation for
the one-particle distribution function $f\left( x,v,t\right) $. A precise
definition of this function will be given later. We will assume in all the
following that the velocities of the particles are of order one. This can
always be assumed by means of a suitable change in the unit of time. Notice
that the microscopic variables are denoted as $X,V,\tau$ while the
macroscopic variables, which will be defined in detail later, will be
denoted as $x,v,t.$ We assume that for a typical particle, $\left\vert
V_{k}\right\vert $ and $\left\vert v_{k}\right\vert $ are of order one and
also that the typical microscopic distance between two particles is of order
one, i.e. $|X_j-X_k|\sim 1 $.

\bigskip

We now describe some families of problems which are simpler than (\ref{S1E1}%
), but that will allow to approximate the dynamics of (\ref{S1E1}) in a
suitable asymptotic limit. The common feature of these models is that they
describe the dynamics of a tracer particle that moves in a random medium.
More precisely, we are interested in the evolution of a particle
characterized by its position and velocity $\left( X\left( \tau\right)
,V\left( \tau\right) \right) $ whose dynamics is given by the system of
equations:%
\begin{equation}
\frac{dX}{d\tau}=V\ \ ,\ \ \frac{dV}{d\tau}=-\Lambda_{\varepsilon}\left(
V\right) +F_{\varepsilon}\left( X,\tau;\omega\right)  \label{S4E8}
\end{equation}
where $F_{\varepsilon}$ is a random force field defined for $\omega$ in a
suitable probability space $\Omega$ and $\Lambda_{\varepsilon}\left(
V\right) $ is a function which can be thought of as a
friction term depending only on the particle velocity $V.$ Notice that
we will not require that $\Lambda_{\varepsilon}\left( V\right) $ and $V$ are
necessarily parallel.

We will assume that the initial positions and velocities $\left\{ \left(
X_{j},V_{j}\right) :j\in S\right\} $ in~\eqref{S1E1} are chosen according to
some probability distribution which is spatially homogeneous and with a
distribution of velocities $g=g\left( v\right) .$ In order to make precise
the connection between (\ref{S1E1}) and~(\ref{S4E8}) we must choose the
random force field $F_{\varepsilon}$ and the friction term $%
\Lambda_{\varepsilon }\left( V\right) $ as functionals of $g,$ i.e.:%
\begin{equation}
\Lambda_{\varepsilon}\left( V\right) =\Lambda_{\varepsilon}\left( V;g\right)
\ \ ,\ \ \ F_{\varepsilon}\left( X,\tau;\omega\right) =F_{\varepsilon}\left(
X,\tau;\omega;g\right).  \label{S5E4}
\end{equation}
In most of the paper we will restrict ourselves to the
study of spatially homogeneous particle distributions. Therefore, we will
assume that the random field $F_{\varepsilon }\left( X,\tau ;\omega \right) $
is invariant under space translations. Some examples of non-homogeneous
particle distributions will be discussed in Section \ref{Nonhomog}. In this
case, in order to obtain consistent kinetic limits we need to assume that
the length scale of the inhomogeneities is comparable to the mean free path
of the system.

\bigskip

A second type of dynamical systems which we will consider in this paper and
that might be used to approximate the dynamics of the system (\ref{S1E1})
are the so-called Rayleigh gases. These systems give the evolution of a
tagged particle in the force field generated by a countable set $S$ of
infinitely many particles (scatterers), each of them is the center of a
potential field. Moreover, it is assumed that the tagged
particle and the scatterers interact by means of the usual Newton's laws,
but that scatterers do not interact among themselves. Suppose that we
denote the position and velocity of the tagged particle as $\left(
X,V\right) $ and the positions and velocities of the scatterers as $\left\{
\left( Y_{k},W_{k}\right) \right\} _{k\in S}.$ We will take a system of
units in which the mass of the tagged particle is $1$ and we will restrict
ourselves for the moment to the case in which the mass of all the scatterers and the
tagged particle is the same. Then, the set of equations describing the
dynamics of a Rayleigh gas is:%
\begin{align}
\frac{dX}{d\tau }& =V\ \ ,\ \ \frac{dV}{d\tau }=-\sum_{j\in S}\nabla \Phi
_{\varepsilon }\left( X-Y_{j}\right),  \notag \\
\frac{dY_{k}}{d\tau }& =W_{k}\ \ ,\ \ \frac{dW_{k}}{d\tau }=-\nabla \Phi
_{\varepsilon }\left( Y_{k}-X\right) \ \ ,\ \ k\in S , \label{S4E9}
\end{align}%
where $\Phi _{\varepsilon }$ is the interaction potential
describing the interaction between the tagged particle and each of the
scatterers.

Note that from the physical point of view, this Rayleigh gas system describes the dynamics of a tracer particle moving in a
background of particles for which their mean free path is much larger than
the mean free path of the tracer particle. Rayleigh gases have been
extensively studied (cf.~\cite{BLLS}, \cite{BGS}, \cite{S1} and \cite{S2}).  
We remark that we call Rayleigh gas the system that is denoted as ideal
Rayleigh gas in \cite{S2} (see also \cite{NWL19}).

We can characterize a Rayleigh gas by means of a random point process in the
phase space $\mathbb{R}^{3}\times \mathbb{R}^{3}.$ A class of measures which
are often used in kinetic theory are the generalized Poisson measures.
These measures are uniquely characterized by a (typically nonfinite) measure $%
g\in \mathcal{M}_{+}\left( \mathbb{R}^{3}\times \mathbb{R}^{3}\right) $
defined in the phase space $\mathbb{R}^{3}\times \mathbb{R}^{3}.$ Then, the
corresponding probability measure is determined uniquely assuming that the
particles are independently distributed and that the average number of
particles in a Borel set $A\subset \mathbb{R}^{3}\times \mathbb{R}^{3}$ is
given by:%
\begin{equation}
\EE\left[\vert {j: (X_j,V_j) \in A}\vert\right]=\int_{A}g\left( dX,dV\right) .  \label{GPoissonMeas}
\end{equation}

It will be convenient to consider measures $g$ yielding bounded particles
densities in the space of particle positions. These measures satisfy: 

\begin{equation}
\rho\in L^{\infty}\left( \mathbb{R}^{3}\right),  \label{boundDens}
\end{equation}
where $\rho$ is defined by 
\begin{equation}
\rho\left( X\right) =\int_{\mathbb{R}^{3}}g\left( X,dV\right).
\label{DensDef}
\end{equation}

\bigskip

We can define a random evolution for the tagged particle $\left( X,V\right) $
assuming that the initial distribution of scatterers in the phase space is
given by a probability distribution. We will restrict ourselves to the case
in which the initial distribution of scatterers $\left\{ \left( Y_{k}\left(
0\right) ,W_{k}\left( 0\right) \right) \right\} _{k\in S}$ is determined
using a generalized Poisson probability distribution associated to a measure 
$g$ in the phase space.

Notice that the problems (\ref{S4E9}) and (\ref{S4E8}) are not equivalent, since the tagged particle modifies the force field  $F_{\varepsilon }\left(X,\tau;\omega \right)=-\sum_{j\in S}\nabla \Phi
_{\varepsilon }\left( X(\tau)-Y_{j}(\tau)\right) $ in (\ref{S4E9}), while the force field is independent of $(X,V)$ in (\ref{S4E8}). However, it turns out that, at least in some scaling limits, the two problems can be translated into one another by an appropriate choice of the friction term $\Lambda_{\varepsilon}\left(V\right)$. In these cases, it is possible to
approximate the dynamics given by (\ref{S4E9}) using the much simpler
dynamics (\ref{S4E8}).

%
%

\bigskip

In this paper, as well as in the companion paper \cite{NVW}, we will introduce a formalism which allows to derive kinetic equations for many particle systems in the case of short-range and long-range interactions. We will mainly focus on the long-range case since the short-range case has been extensively studied in the mathematical physics community. 
 We will provide the main ideas of the derivation of these  kinetic equations, as well as the precise scaling limits, in this paper. We will postpone to \cite{NVW}  some technical arguments related to the computation of the so called dielectric function. This is a function that allows to approximate the collective behaviour of the system which is due to the long-range interactions.

%
Most of the long-range interaction potentials we consider in these two papers are weak interaction potentials with a range much larger than the
typical distance between particles. In this situation, under suitable
assumptions, there exists a macroscopic time scale $T_{\varepsilon }$ which
is much larger than the time required for a particle to travel the typical
distance between particles. The time $T_{\varepsilon }$ is the typical time
in which the velocity of one particle changes by an amount of order one. We
can define a macroscopic time variable $t=\frac{\tau }{T_{\varepsilon }}.$
It turns out that in some scaling limits (see Section \ref{sec:ManyPartFrict}) we
can approximate for small but macroscopic times, the dynamics (\ref{S1E1})
by means of at least one of the dynamics (\ref{S4E8}), (\ref{S4E9})
depending on the form of the interaction potentials.


In order to define the one-particle distribution function for the tracer particle, we consider an ensemble $\left\{ \left( 
X_{j},V_{j}\right) \right\}_{j\in S} $ of tracer particles, chosen according to to a
generalized Poisson distribution $\mathbb{\tilde{P}}_{0}$ with intensity 
measure $f_{0}\in\mathcal{M}_{+}\left( 
\mathbb{R}^{3}\times\mathbb{R}^{3}\right) $ (cf.~(\ref{GPoissonMeas})).
 We will assume also that
the density $\rho_{0}$ defined by means of $\rho_{0}\left( X\right) =\int_{%
	\mathbb{R}^{3}}f_{0}\left( X,dV\right) $ satisfies $\rho_{0}\in
L^{\infty}\left( \mathbb{R}^{3}\right) $ (cf.~(\ref{boundDens}), (\ref%
{DensDef})). 
In both the cases (\ref{S4E8}), (\ref{S4E9}), we then also consider an independent family of realizations $F_\eps(\cdot ;\omega)$, or $(Y_k(0;\omega),W_k(0;\omega))_{k\in S}$ for all scaling parameters $\eps>0$ and $j\in S$. 
The one-particle distribution function $f_{\varepsilon}\left( X,V;\tau\right) $  for $\tau>0$ is then defined as the intensity measure of the ensemble $\left\{ \left( 
X_{j}(\tau),V_{j}(\tau)\right) \right\} $ determined by either of the evolutions (\ref{S4E8}), (\ref{S4E9}). Notice that due to the assumption of independence, this ensemble is again defined by a generalized Poisson measure $\mathbb{\tilde{P}}^\varepsilon_{\tau}$, hence $f_{\varepsilon}\left( X,V;\tau\right) $ is well-defined.

It turns out that for suitable choices of the random force fields $%
F_{\varepsilon }$ and friction coefficients $\Lambda _{\varepsilon }$ in (%
\ref{S4E8}) or the interaction potentials $\Phi _{\varepsilon }$ in (\ref%
{S4E9}) we can approximate the evolution of the functions $f_{\varepsilon
}\left( X,V;\tau \right) $ as $\varepsilon \rightarrow 0$ by means of a
function $f$ which satisfies a Markovian integro-differential equation. More
precisely, the following limit exists in the sense of measures:%
\begin{equation}
f\left( x,v,t\right) =\lim_{\varepsilon \rightarrow 0}f_{\varepsilon }\left(
T_{\varepsilon }x,v;T_{\varepsilon }t\right)  \label{S5E3}
\end{equation}%
where $T_{\varepsilon }$ is the macroscopic time scale defined above and it
satisfies $\lim_{\varepsilon \rightarrow 0}T_{\varepsilon }=\infty.$ We will
denote the variables $x$ and $t$ in (\ref{S5E3}) as macroscopic space and
macroscopic time respectively. 
Moreover, the measure $f$ solves a linear equation with the form:%
\begin{equation}
\partial _{t}f+v\cdot \partial _{x}f=K\left[ f(t,x,\cdot)\right](v) \ \ ,\ \
f\left( \cdot ,\cdot ,0\right) =f_0(\cdot )  \label{KinLimTag}
\end{equation}%
where $K\left[ \cdot \right] $ is an integro-differential operator,
depending on the type of interactions acting only on the velocity variable
of $f$ for each fixed $x$. The precise form of the operator $K\left[ \cdot \right] $, for different types of
particle interactions, will be described in Section \ref{sec:KinEq}. 

Equation (\ref{KinLimTag}) contains all the information about the kinetic
regime which gives the dynamics as $\varepsilon\rightarrow0$ for particles
evolving according to (\ref{S4E8}) or (\ref{S4E9}). If we consider random
force fields depending on a distribution $g$ of particle velocities as in (%
\ref{S5E4}) we would obtain a collision operator $K\left[ \cdot\right] $
depending also on $g.$ Then, the corresponding kinetic equation would take
the form:%
\begin{equation}
\partial_{t}f+v\cdot\nabla_{x}f=K\left[ f;g\right]  \label{KinLimNonHom}
\end{equation}
where the operator $f\rightarrow K\left[ f;g\right] $ acts only on the
variables $v$ of $f$ at each point $x\in\mathbb{R}^{3}.$ Then, in the
homogeneous case (\ref{KinLimNonHom}) reduces to:%
\begin{equation}
\partial_{t}f=K\left[ f;g\right] .  \label{KinLin}
\end{equation}

We can now state the relation between the kinetic limit for the dynamic of
the many particle system given by (\ref{S1E1}) and the kinetic limits for (%
\ref{S4E8}) and (\ref{S4E9}) described above. Suppose that we define a
generalized Poisson measure $\mathbb{P}_{0}$ given by a density $f_{0}\in 
\mathcal{M}_{+}\left( \mathbb{R}^{3}\times \mathbb{R}^{3}\right) $ as above.
We consider then the evolution of the particle configurations by means of (%
\ref{S1E1}). Assuming the dynamics \eqref{S1E1} is well-defined with $%
\mathbb{P}_{0}-$probability one, we can define the associated evolution
mapping $U^\tau$: 
\begin{align*}
U^\tau_{\varepsilon} ((X_k,V_k)_{k\in S}) = (X_k(\tau),V_k(\tau)),
\end{align*}
with associated inverse $U^{-\tau}$. This allow to define a new measure $%
\mathbb{P}_{t}^{\varepsilon }$ for each positive time by means of 
\begin{align*}
\mathbb{P}_{\tau }^{\varepsilon }= \mathbb{P}_{0} \circ
U^{-\tau}_{\varepsilon} .
\end{align*}
In the case of the evolution given by (\ref{S1E1}) the new probability
measure is not a generalized Poisson measure for positive times anymore. The
difference between the probability measure $\mathbb{P}_{\tau }^{\varepsilon
} $ and a generalized Poisson measure is usually measured using the
correlation functions appearing in the BBGKY hierarchies (\cite{Bo}, \cite{Ce}, \cite{S}), 
although we will not use this approach in most of this
paper (except by a short discussion in Section \ref{correlations}). However,
in the kinetic regime, the probability measures $\mathbb{P}_{\tau
}^{\varepsilon }$ converge as $\varepsilon \rightarrow 0$ to a generalized
Poisson measure which can be characterized by a measure $f$ in the phase
space. We claim that in the kinetic regime the evolution of the measure can
be computed by means of the nonlinear equation:%
\begin{equation}
\partial _{t}f+v\cdot \partial _{x}f=Q\left[ f\right],
\label{KinNonLinNonHom}
\end{equation}%
where $Q$ is given by 
\begin{equation}
Q\left[ f\right] =K\left[ f;f\right]  \label{ClosRel}
\end{equation}%
and $K\left[ \cdot ;g\right] $ is the collision operator for the reduced
model in \eqref{KinLimNonHom}, associated to the density $g$. In the case of
homogeneous particle distributions (\ref{KinNonLinNonHom}) reduces to:%
\begin{equation}
\partial _{t}f=Q\left[ f\right] .  \label{KinNonLin}
\end{equation}

The rationale behind the closure assumption is the following. Suppose that
at a given time $t$ the distribution of particles in the phase space is
characterized by the measure $f\left( \cdot ,t\right) .$ We claim that the
evolution of an individual particle can be approximated, at least for small
macroscopic times for which $f$ is approximately constant, by means of one
of the dynamics (\ref{S4E8}) or (\ref{S4E9}) with a distribution of
particles given by $g=f\left( \cdot ,t\right) .$ Therefore, due to (\ref%
{KinLimNonHom}), the distribution of particles at time $t+h$ can be expected
to be given by:%
\begin{equation}
f\left( x,v,t+h\right) =f\left( x,v,t\right) +h\left[ -v\cdot \partial
_{x}f  +K\left[ f;f \right] \right]
\left( x,v,t\right)  \label{A1}
\end{equation}%
and taking the limit $h\rightarrow 0$ we obtain (\ref{KinNonLinNonHom}), (%
\ref{ClosRel}).

The operator $K$ in (\ref{A1}) is the corresponding one to the kinetic limit
associated to (\ref{S4E8}) or (\ref{S4E9}), depending on the type of
interaction under consideration. On the other hand, as we indicated above,
the dynamics of the Rayleigh gases described by the equations (\ref{S4E9}),
can be approximated in the case of long-range interactions and in the
kinetic limit by means of the dynamics of a tagged particle with friction
moving in a random force field (cf.~(\ref{S4E8})). 
This will allow to derive in all the cases the kinetic equations for long-range potentials by identifying the corresponding kinetic limits for (\ref{S4E8}), a task much simpler than directly deriving the kinetic limits from the original system of equations (\ref{S1E1}). 
One of the main questions that needs to be answered in order to fulfill this
program is to obtain the formulas for the random force field $F_{\varepsilon
}$ and the friction coefficient $\Lambda _{\varepsilon }$ appearing in (\ref%
{S4E8}) from the interactions $\Phi _{\varepsilon }$ for the Rayleigh gas (%
\ref{S4E9}). This will be made in Section~\ref{sec:ManyPartFrict}, and some details are postponed to the companion paper \cite{NVW}.

\bigskip

It is worth to remark that the approximation of the dynamics of a particle
moving in a ``medium", which consists of a large number of particles, by means of
one equation with the form (\ref{S4E8}) is extensively used in Statistical
Physics, not only in situations in which the dynamics of the system can be
approximated by a kinetic equation. A well known example is the description
of the Brownian motion of a particle in a viscous fluid (cf.~\cite{Ein05}). The dynamics of the
Brownian particle can be approximated using a Langevin equation, which
contains a friction term acting on the Brownian particle, and a random term
(white noise). The main difference between this problem and the problems
considered in this paper is that to find the connection between the
microscopic dynamics and the macroscopic coefficient characterizing the
properties of the medium (in this case the friction coefficient), cannot be
made in a manner so explicit in the case of a viscous fluid (described by
Stokes equations) as in the cases in which the dynamics of the medium can be
approximated by a kinetic equation. It is well known that there is a general
formula connecting the properties of the noise term in (\ref{S4E8}) with the
friction coefficient, known as the fluctuation-dissipation Theorem (cf.~\cite{Ku66}). 
The fluctuation-dissipation theorem can be expected to hold for
systems in which there exists a mechanism driving the distribution of the
scatterers to an equilibrium (i.e. not for Rayleigh gases). This connection
between noise and friction is due to the fact that the equilibrium
distribution at a given temperature is the Gibbs distribution, which in the
kinetic regime reduces to Maxwellian distributions for both the tagged
particle and the scatterers. The variances of the velocities are related
through the principle of equipartition of energy which holds at equilibrium.
The equation describing the evolution of the tagged particle is a
Fokker-Planck equation containing a friction term and a diffusive term in the
space of velocities, which must be related in order to yield the decided
value of the variance.

\bigskip

Three classes of kinetic equations with the form (\ref{KinNonLinNonHom}), (%
\ref{ClosRel}) have been used extensively in the physics literature to
approximate the dynamics of systems described by means of (\ref{S1E1}),
namely the Boltzmann equation, the Landau equation and the Balescu-Lenard
equation. The kinetic approximations are valid if the characteristic
potential energy of each particle does not exceed its kinetic energy. This
means that the particle trajectories are nearly rectilinear over lengths of
the order of the typical distance between particles due to the weakness of
the interactions. However, this might happen in two different ways. First we
can have strong interactions, with a range much shorter than the typical
particle distance. In this case, the particles of the system interact
rarely, but when they do, they experience velocity deflections of order one.
The resulting kinetic equation is the Boltzmann equation. The second
possibility is to have weak interaction potentials between particles with a
range comparable larger or equal than the typical particle distance. In this
case the deviations of the trajectory from a rectilinear path are due to the
accumulation of many small random deflections which are due to the
interaction with many different particles. In this case the resulting
kinetic equation is the Landau or the Balescu-Lenard equation. The
difference between Landau and Balescu-Lenard stems from the different forms
of the random force field $F_{\varepsilon }$ and the friction coefficient $%
\Lambda _{\varepsilon }$ in (\ref{S4E8}). Details about the computation of $%
F_{\varepsilon }$ and $\Lambda _{\varepsilon }$ from the particle
interactions will be given in \cite{NVW}. Here we just
indicate that in the cases in which the Landau equation is applicable, $%
F_{\varepsilon }$ and $\Lambda _{\varepsilon }$ are computed assuming that
scatterers move in straight lines. In the case in which the limit kinetic
equation is Balescu-Lenard, we must take into account in the computation of $%
F_{\varepsilon }$ and $\Lambda _{\varepsilon }$ the interactions between the
background particles. The alternative depends on the type of
interactions taking place between the particles in (\ref{S1E1}). In the
cases in which the Balescu-Lenard equation is applicable, the mutual
interactions between the background particles can be described using a single function
which is usually termed as dielectric function. This function and its main
properties will be discussed in \cite{NVW}. Notice that the
Landau equation is just a particular case of the Balescu-Lenard equation in
which the dielectric function is constant with value one.

It is worth to mention that the Landau equation can arise as a kinetic limit
of (\ref{S1E1}) for some short range potentials. The typical situation in
which this can happen is the so-called grazing collisions limit which is
characterized by weak interaction potentials with a range much shorter or
comparable to the typical particle distance. In such a system the changes
in the velocity of one particle are due to the sum of the effects of many
pairwise independent weak collisions, each of them yielding small
deflections. We refer for instance to \cite{AlVi, DV1, DV2, Gou}.

\bigskip

In principle, it is not possible to approximate (\ref{S1E1}) by means of the
dynamics (\ref{S4E8}) in the case in which the resulting kinetic equation is
a Boltzmann equation. This is due to the fact that the friction term $%
\Lambda _{\varepsilon }$ in (\ref{S4E8}) is due to the long range
interaction of many scatterers acting on the tagged particle, while in the
cases in which the kinetic description is given by the Boltzmann equation the interactions
between particles are due only to pairwise collisions. Therefore, the
Boltzmann kinetic regime can be obtained only approximating (\ref{S1E1}) by
means of a Rayleigh gas dynamics as in (\ref{S4E9}). However, it will be
seen in Subsection \ref{BotzInel} that the dynamics of a tagged particle in
such a Rayleigh gas can be replaced in the case of hard sphere potentials by
a new system with inelastic collisions which in the limit $\varepsilon
\rightarrow 0$ yields the same kinetic dynamics as the original Rayleigh
gas. The inelastic collisions can be thought as the analogous of the
friction coefficient $\Lambda _{\varepsilon }$ in the Boltzmann kinetic
regime.

\bigskip

The derivation of the three types of kinetic equations discussed above has
been extensively considered in the physical literature. The Landau and the
Balescu-Lenard equations have been considered mostly in the plasma physics
literature, in the particular case in which the particles interact by means
of Coulomb potentials. We will consider the Coulombian case in \cite{NVW}, as well as more general interaction potentials.

The connection between the Balescu-Lenard equation and the dynamics of
tagged particles in effective media has been pointed out by several authors
(cf.~\cite{Lan1, Lan2, LL2, Pi87}).  
The roots of some of the ideas explained above in order to obtain the
kinetic limit of systems of particles interacting by means of long range
potentials can be found in the seminal paper by Bogoliubov (cf.~\cite{Bo}), in
which the fluctuations of the particle distributions are described using
BBGKY hierarchies (see also~\cite{Ba1, Ba2, Le}). {We would also like to point out a detailed review in the physics literature  on the emergence of kinetic equations from long-range interacting systems where issues closely related to the ones in this paper have been considered  (cf.~\cite{Ch1,Ch2,Ch3})}.

\bigskip

The kinetic limits which describe the dynamics of a tagged particle moving
among a set of fixed scatterers, interacting by means of long-range potentials, have been considered in \cite{NSV}. Systems in which a tracer particle interacts with scatterers which are not affected
by the tracer particle are usually called Lorentz gases (cf.~\cite{Lo}). The Lorentz gas can
be obtained as a formal limit of a Rayleigh gas in which the mass of the
scatterers is much larger than the mass of the tagged particle.

In this case the distribution function $f( x,v,t)$ yields the probability
of finding the tagged particle at a position of the phase space. The
evolution of this function is described in the resulting kinetic limits
studied in \cite{NSV} either by a linear Boltzmann equation or by a linear
Landau equation. Rigorous results on the derivation of these linear kinetic equations have been provided in cases of compactly supported potentials. We refer to \cite{DGL, G, KP,S3} for first results in this direction and to \cite{BNP, BNPP, DR, DP, LT, MN} for related results and subsequent developments.  

The main conclusion obtained in \cite{NSV} for Lorentz gases are the
following. The description of $f\left( x,v,t\right) $ by means of a kinetic
equation is possible only if the kinetic energy of a typical particle is
much larger than its potential energy. As indicated above this can happen in
two different ways, namely because the range of the potentials is much
shorter than the average particle distance, or because the interaction
potentials are very weak and the tagged particle deflections are due to the
addition of many small independent deflections due to different particles.
It has been seen in \cite{NSV} that in order to decide if the limit kinetic
equation is a Boltzmann or a Landau equation it is convenient to compute two
different time scales $T_{BG}$ and $T_{L}$ which are denoted as
Boltzmann-Grad and Landau time scales respectively. The time $T_{BG}$ is the
typical time in which the tagged particle arrives sufficiently close to one
scatterer to experience a deflection of its velocity comparable to the
velocity itself. The time scale $T_{L}$ is the characteristic time in which
the velocity of the tagged particle experiences a change comparable to it
due to the accumulation of small random deflections.

It was proved in \cite{NSV} that the dynamics of a tagged particle can be
described by means of a linear Boltzmann equation if $1\ll T_{BG}\ll T_{L}.$
On the other hand, the description by means of a Landau equation is possible
if $1\ll T_{L}\ll T_{BG}.$ The time scale in which the kinetic evolution
takes place is $T_{\varepsilon}=\min\left\{ T_{BG},T_{L}\right\} .$ A second
condition which the interactions must satisfy is that they must become
independent on distances of order $T_{\varepsilon}$ (assuming that the
characteristic velocity of the tagged particle is of order one). Actually,
it has been shown in \cite{NSV} that for some choices of interaction
potentials the deflections of the tagged particles at times of the same
order as $T_{\varepsilon}=T_{L}$ have correlations of order one and the
resulting limit dynamics cannot be described by a kinetic equation.

It turns out that for interacting particle systems (cf.~\eqref{S1E1}), as well
as for Rayleigh gases, it is possible to define in an analogous manner the
characteristic time scales $T_{BG}$ and $T_{L}$. The time scales $T_{BG}$, $T_{L}$ allow us to determine whether the particle system can be
described as a kinetic limit, and whether the resulting equation is a
Boltzmann equation or a Landau/Balescu-Lenard equation. If we assume that
the particle velocities are of order one the mean free path is $%
T_{\varepsilon }.$ The two conditions required to obtain a kinetic equation
in the limit $\varepsilon \rightarrow 0$ are the same as those obtained in
the case of fixed scatterers, namely 
\begin{itemize}
	\item the mean free path is much larger
	than the typical particle distance (i.e. $T_{\varepsilon }\gg 1$) 
	\item the deflections of the particle velocities for particles at distances larger
	than the mean free path become uncorrelated as $\varepsilon \rightarrow 0$.
\end{itemize}
The characteristic time scales $T_{\ep}$ that will be introduced in Sections \ref{ss:KinEqL}, \ref{ss:KinEqR}, \ref{ss:KinEqI} will be computed in detail in the companion paper \cite{NVW}. 

\bigskip

\subsection{Conjectures on the kinetic equations for scaling limits of interacting particle systems in a spatially homogeneous setting}

As explained above, we will take as unit length the typical interparticle distance, i.e. $d=1$, for most of this paper. We then compute the kinetic timescale, i.e. the time scale on which the normalized  one-particle function $f_1^\eps$ changes an amount of order one. This choice of units has the advantage that we do not need to know the correct timescale before starting our analysis. 

In kinetic theory, it is however customary to formulate scaling limits in the macroscopic length. In this section, we will summarize our analysis in terms of scaling limits of interacting particle systems. Here $N$ will be the average number of particles per macroscopic unit volume, and $r=\ep$ the radius of interaction.

For smooth decaying potentials, we consider the following parametric family of scaling limits $\eps \rightarrow 0$:
\begin{align}\label{scaling:soft}
\Phi_\eps(x)= \eps^\alpha \Phi(x/\ep), \quad N = \eps^{-\beta} . 
\end{align}
\begin{remark}
	Notice that we assume the radius of interaction $\ep\rightarrow 0$ vanishes in the limit. For a radius of interaction of order one, i.e. $\Phi_\eps(x)= \eps^\alpha \Phi(x)$, the Vlasov equation can be obtained in the mean field scaling $\alpha=\beta$.
\end{remark}
\begin{conjecture}[Smooth decaying potentials]
	In the scaling \eqref{scaling:soft}, the normalized one-particle marginal $f^\eps_1$ converges to a solution of a kinetic equation if $\alpha\in [0,1]$ and $\beta>0$ satisfy
	\begin{align}
	\beta = 2(1+\alpha).
	\end{align}
	The kinetic equation then depends on $\alpha \in [0,1]$ as follows:
	\begin{enumerate}
		\item	Boltzmann equation for $\alpha= 0$
		\item 	Landau equation for $\alpha\in (0,1)$
		\item 	Balescu-Lenard equation for $\alpha=1$.
	\end{enumerate}
\end{conjecture}
In the case of scale-invariant potentials of the form
\begin{align}\label{rPot}
\Phi(x) = 1/|x|^s, \quad s>\frac12,
\end{align}
scaling limits can be parametrized as
\begin{align}\label{scaling:rPot}
\Phi_\eps(x) = \eps \Phi(x), \quad N=\eps^{-\gamma},
\end{align}
with $\gamma>0$ to be chosen. In this case, our conjecture is the following.
\begin{conjecture}[Scale invariant potentials]
	Let $\Phi$ be a scale invariant potential of the form \eqref{rPot}. Then, 
	\begin{itemize}
	\item if $s\in (\frac12,1)$, the one-particle marginal $f_1^\eps$ converges to a solution of a kinetic equation if 
	\begin{align}
	\gamma = 4/(1+s),
	\end{align}
	and the resulting limit equation is the Balescu-Lenard equation. 
	\item If $s=1$, on the logarithmic time scale defined by $\bar{t}=\vert \log\ep \vert  t$, $f_1^{\eps}$ converges to a solution of the Landau equation.   
	\item If $s>1$, then we obtain a kinetic limit for 
	\begin{align}
	\gamma = 2/s,
	\end{align}
	and the resulting kinetic equation is the Boltzmann equation.
	\end{itemize}
\end{conjecture}
\begin{remark}
	For scaling limits which lead to the Boltzmann equation, the appropriate distribution of initial data for the particle system are not Poisson distributions. Indeed, the initial probability measure is modified in order to avoid too many particle overlappings  (see for instance \cite{GST}, \cite{PSS}). Since our focus in this work is on systems governed by long-range interactions, we will not discuss this in detail here. 
\end{remark}

\subsection{Generalization to spatially inhomogeneous systems. \label{Nonhomog}} 
 
 We have considered in the previous Sections only spatially homogeneous
particle distributions. We can apply similar methods to study some classes
of non spatially homogeneous situations. We will not consider the details in
this paper, but we just indicate under which assumptions we can obtain
kinetic equations analogous to the ones obtained in the homogeneous case.

The specific scaling limit in which we can obtain a non spatially homogeneous
kinetic equation is the following. Suppose that we consider a set of
particles in the phase space $\mathbb{R}^{3}\times \mathbb{R}^{3}$ by means
of a non homogeneous Poisson distribution characterized by the rate:%
\begin{equation*}
f_{0}\left( \frac{x}{L_{h}},v\right),\qquad L_{h}\gg 1.
\end{equation*}%
We assume that the average distance between particles is one and that the
characteristic velocity of the particles is of order one too. Suppose that
the particle configuration evolves according to the system of equations (\ref%
{S1E1}). Choosing $L_{h}$ as the mean free path we obtain the version of the
kinetic equation containing the transport term $v\cdot \nabla_{x}$ which
does not vanish for nonhomogeneous solutions.

More precisely, in the case of Rayleigh Gases we obtain the following family
of linear kinetic equations 
\begin{equation}  \label{eq:nonhomeq1}
\partial _{t}f+v\cdot \nabla_{x}f=Q(g,f) \ , \quad f=f(x,v,t)
\end{equation}
where the linear collision operator $Q=Q_{B}$ with $Q_{B}$ given as in %
\eqref{eq:opQ} in the case of the Rayleigh-Boltzmann equation, and $Q=Q_{L}$
with $Q_{L}$ given as in \eqref{LandRayEq} in the case of the
Rayleigh-Landau equation. \smallskip

In the case of interacting particle systems we obtain the following family
of nonlinear kinetic equations 
\begin{equation}  \label{eq:nonhomeq2}
\partial _{t}f+v\cdot \nabla_{x}f=Q(f,f) \ , \quad f=f(x,v,t)
\end{equation}
where the collision operator $Q=Q_{B}$ with $Q_{B}$ given as in %
\eqref{eq:boltzman} in the case of the Boltzmann equation, $Q=Q_{{L}}$ with $%
Q_{{L}}$ given as in \eqref{eq:landau} in the case of the Landau equation
and $Q=Q_{BL}$ with $Q=Q_{BL}$ given as in \eqref{eq:BalescuLenard} in the
case of the Balescu-Lenard equation. 

The non homogeneous kinetic equations are meaningful only if $L_{h}$ is
comparable or much larger than the mean free path. If $L_{h}$ is much larger
than the mean free path we would obtain that the kinetic equations would be
given approximately by homogeneous equations, whose solutions converge to
Maxwellian distributions characterized by local temperatures, density and
momentum which would change in the characteristic scale $L_{h}.$ This
situation corresponds to the so-called hydrodynamic limit that we will not discuss in this paper.

\bigskip

\subsection{Notation and structure of the paper.}

We will use the following notation in this paper:%
\begin{equation}
\mathbb{N}_{\ast }=\left\{ 0,1,2,3,....\right\} \ ,\ \ \mathbb{N}=\left\{
1,2,3,....\right\} . \label{SetN}
\end{equation}
For a function or measure $g(x,v)$ on the phase space, we denote the
associated spatial density by $\rho[g]$ by: 
\begin{equation}
\rho [ g]\left( x\right) =\int_{\mathbb{R}^{3}}g\left( x,w\right) dw.
\label{eq:spatdensity}
\end{equation}

The plan of this paper is the following. 
In Section \ref{sec:KinEq} we summarize the
different kinetic equations that can be obtained to approximate the dynamics
of systems with the form (\ref{S1E1}) in suitable asymptotic limits. 
In Section \ref{sec:ManyPartFrict} we show how to approximate the dynamics of both a tagged
particle in a Rayleigh gas (cf.~(\ref{S4E9})) or a system of interacting
particles (cf.~(\ref{S1E1})) by a dynamic of the form \eqref{S4E8}, in cases in
which the interactions take place by means of long-range potentials.  
In this section we also introduce the Boltzmann time scale and the Landau time scale in a manner analogous to \cite{NSV}.  
Section \ref{correlations} compares the approach used in Section \ref{sec:ManyPartFrict} to derive kinetic equations with
the classical approach proposed by Bogoliubov based in BBKGY hierarchies.    
Section \ref{sec:misc} consists of two subsections. In the first one we review some phenomena
associated to the Vlasov equation which play an important role in the kinetic theory of particle systems with long-range interactions. 
In the second subsection we will show that we can derive kinetic equations from many particle systems with hard-sphere interactions analysing the dynamics of a Rayleigh Gas with non-elastic collisions. This approach is reminiscent of the approach used in Section \ref{sec:ManyPartFrict} to obtain nonlinear kinetic equations from the dynamics of a Rayleigh Gas containing a friction term. We conclude the paper with a summary of the results obtained here, which is the content of Section \ref{sec:concl}. 

\section{Kinetic descriptions of many particle systems.}\label{sec:KinEq}

We summarize in this Section the different kinetic equations which
approximate the dynamics of systems of particles described by the equations (%
\ref{S1E1}) as well as the form of the interaction potentials and the
asymptotic limits in which such kinetic approximation is valid.

\bigskip

We recall briefly the features that characterize each of the models at the
level of particles: 
\begin{equation*}
\begin{array}{|c|c|c|cc|}
\hline
\text{Model} & 
\begin{array}{c}
\text{Scatterers influence} \\ 
\text{tagged particle}%
\end{array}
& 
\begin{array}{c}
\text{Tagged particle} \\ 
\text{influences scattereres}%
\end{array}
& 
\begin{array}{c}
\text{Scatterers influence} \\ 
\text{other scatterers}%
\end{array}
&  \\ \hline
&  &  &  &  \\ 
\text{Lorentz gas} & \text{Yes} & \text{No} & \text{No} &  \\ \hline
&  &  &  &  \\ 
\text{Rayleigh gas} & \text{Yes} & \text{Yes} & \text{No} &  \\ \hline
\begin{array}{c}
\text{Interacting particle} \\ 
\text{system}%
\end{array}
& \text{Yes} & \text{Yes} & \text{Yes} &  \\ \hline
\end{array}%
\end{equation*}

\bigskip

\begin{center}
$%
\begin{array}{|c|c|cc|}
\hline
\text{Model} & 
\begin{array}{c}
\text{Momentum conserved} \\ 
\text{in the collisions}%
\end{array}
& 
\begin{array}{c}
\text{Energy conserved} \\ 
\text{in the collisions}%
\end{array}
&  \\ \hline
&  &  &  \\ 
\text{Lorentz gas} & \text{No} & 
\begin{array}{c}
\text{No\ (moving scatterers)} \\ 
\text{Yes (fixed scatterers)}%
\end{array}
&  \\ \hline
&  &  &  \\ 
\text{Rayleigh gas} & \text{Yes} & \text{Yes} &  \\ \hline
\begin{array}{c}
\text{Interacting particle} \\ 
\text{system}%
\end{array}
& \text{Yes} & \text{Yes} &  \\ \hline
\end{array}%
$
\end{center}

\medskip

\subsection{Evolution of a tagged particle in a Lorentz gas.}\label{ss:KinEqL}

\bigskip

In the next three Subsections we describe the kinetic equations which
describe the behaviour of Lorentz and Rayleigh gases as well as interacting
particle systems in suitable kinetic limits and for suitable classes of
potentials. We restrict ourselves here for simplicity to the case of
spatially homogeneous systems.

The dynamics of a tagged particle in a Lorentz gas is described by the
following set of evolution equations: 
\begin{align}
\frac{dX}{d\tau }& =V\ \ ,\ \ \frac{dV}{d\tau }=-\sum_{j\in S}\nabla \Phi
_{\varepsilon }\left( X-Y_{j}\right)  \notag \\
\frac{dY_{k}}{d\tau }& =W_{k}\ \ ,\ \ \frac{dW_{k}}{d\tau }=0\ \ ,\ \ k\in S \ .
\label{NewtLor}
\end{align}

We observe that the Lorentz gas can be thought as the interaction of tagged
particles with a random force field (static or moving). 
\bigskip

As it has been proved in \cite{NSV} it is possible to show that different
linear kinetic evolution equations arise, for a given scaling, depending on
the decay $s$ as well as on the singularities of the interaction potential,
assuming that all the scatterers are at rest, i.e. $W_{k}=0$ for all $k\in
S. $ Using similar methods, it is possible to obtain the following equations
describing the evolution of the distribution of velocities $f\left(
v,t\right) $ which characterizes the tagged particle. We assume in all the
following that the typical interparticle distance between scatterers is $%
d=1. $ We will assume also that the mean free path is much larger than the
interparticle distance, i.e. $\ell _{\varepsilon }>>d$. We then have the
following possibilities.

\begin{itemize}
\item[(1)] For potentials with the form $\Phi _{\varepsilon }\left( Y\right)
=\Phi \left( \frac{\left\vert Y\right\vert }{\varepsilon }\right) ,$ with $%
\varepsilon \rightarrow 0$ and $ \Phi \left( r\right) $ decreasing faster than $%
\frac{1}{r^{s}},$ with $s >1$, the kinetic time scale is $%
T_{\varepsilon }=\frac{1}{\varepsilon ^{2}}$ and the resulting equation is
the linear Boltzmann (with an additional averaging due to the possible
distribution of velocities of the scatterers): %
\begin{align}
\partial _{t}f\left( v,t\right) &=\mathcal{L}_{B}(f)(v,t)  \notag \\
\mathcal{L}_{B}(f)(v)&=\int_{\mathbb{R}^{3}}dv_{\ast }\int_{S^{2}}d\omega \ 
\tilde{B}\left( \omega \cdot \left( v-v_{\ast }\right) ,\left\vert v-v_{\ast
}\right\vert \right) g\left( v_{\ast }\right) \left[ f\left( v^{\prime
}\right) -f\left( v\right) \right]  \label{boltLorMov}
\end{align}%
where 
\begin{equation*}
v^{\prime }=v-2\left( \omega \cdot \left( v-v_{\ast }\right) \right) \omega \ .
\end{equation*}
The cross section $\tilde{B}$ is different from the one computed for the
usual collisions between two particles because we assume that the scatterer
is not affected by the tagged particle. The simplest way of studying this
scattering process is to use a coordinate system moving at the scatterer
speed $v_{\ast }.$ Notice that the solutions of (\ref{boltLorMov}) do not
preserve the energy or momentum for the distribution of tagged particles,
something that could be expected since these quantities are not preserved in
the individual collisions.

\item[(2)] For potentials with the form $\Phi _{\varepsilon }\left( Y\right)
=\varepsilon \Phi \left( \varepsilon ^{s }\left\vert Y\right\vert \right) ,$
with $\varepsilon \rightarrow 0,\ s <\frac{2}{5},\ \Phi \left( r\right) $
decreasing faster than $\frac{1}{r^{s }}$ as $s\rightarrow \infty ,$
with $s >1,$ assuming that the average particle distance is $d=1$ 
and choosing the time scale $T_{\varepsilon }=\varepsilon ^{4s -2}$
which satisfies $T_{\varepsilon }\gg 1$ as $\varepsilon \rightarrow 0$ we
obtain that the distribution of particle velocities of the tagged particle $%
f\left( v,t\right) $ solves the kinetic equation 
\begin{align}
\partial _{t}f\left( v,t\right) & =\mathcal{L}_{L}(f)(v,t)  \notag \\
\mathcal{L}_{L}(f)(v)& =\sum_{j,k}\int_{\mathbb{R}^{3}}dv_{\ast }g\left(
v_{\ast }\right) \partial _{j}\left( \kappa \left( \delta _{j,k}-\frac{%
\left( v-v_{\ast }\right) _{j}\left( v-v_{\ast }\right) _{k}}{\left\vert
v-v_{\ast }\right\vert ^{2}}\right) \partial _{k}f\left( v,t\right) \right)
\label{LandLorMov}
\end{align}%
where $\kappa $ is the diffusion coefficient in the space of particle
velocities, which is given by:%
\begin{equation}
\kappa =\frac{\pi}{4}\frac{1}{|v-v_{\ast }|}\int_{\R^3}dk |\hat{\Phi}%
(|k|)|^{2}|k| \label{def:kappa}
\end{equation}
and $\hat{\Phi}$ is the Fourier transform of $V$, i.e. $\hat{\Phi}(k)=\frac{1}{(2\pi)^{\frac 3 2}}\int_{\R^3} dx \Phi(x) e^{-ik\cdot x}$. Similar equations can be
obtained in the case of potentials $\Phi\left( r\right) $ behaving as $\frac{1}{%
r}$ as $r\rightarrow \infty .$ Nevertheless the formula (\ref{def:kappa})
has to be modified, due to the presence of logarithmic divergences. These
divergences can be compensated by means of a suitable change in the time
scale. We refer to the companion paper \cite{NVW} for details about this
case. A similar situation can be seen in the case of fixed scatterers in
Subsection 3.2 of \cite{NSV}.
\end{itemize}

\bigskip

The solutions of (\ref{LandLorMov}) do not preserve neither the energy or
the momentum of the particle distribution $f\left( v,t\right) .$ It is worth
to remark that the equations obtained in \cite{NSV} in the case of
scatterers at rest can be obtained as special cases of the equations (\ref%
{boltLorMov}), (\ref{LandLorMov}) just taking the distribution $g\left(
v_{\ast }\right) =\delta \left( v_{\ast }\right) .$ Actually, more general
classes of potentials than those described in the point (2), including
potentials $\Phi \left( r\right) $ behaving as Coulombian potentials for large $%
r,$ have been considered in \cite{NSV}. In such cases a logarithmic
correction, termed as Coulombian logarithm, must be included in the time
scale $T_{\varepsilon }.$ The reason for the onset of the Coulombian logarithm is the fact that all the particles colliding with impact parameter $b \in [2^k,2^{k+1}]$, $k\in \N$, give a contribution of the same order of magnitute to the particles deflections.

\bigskip

\subsection{Evolution of a tagged particle in a Rayleigh gas.}\label{ss:KinEqR}

\bigskip

We summarize now the classes of kinetic equations that can be obtained for
Rayleigh gases. We recall that the set of equations describing the dynamics
of a Rayleigh gas is given by \eqref{S4E9}, namely 
\begin{align*}
\frac{dX}{d\tau }& =V\ \ ,\ \ \frac{dV}{d\tau }=-\sum_{j\in S}\nabla \Phi
_{\varepsilon }\left( X-Y_{j}\right) \\
\frac{dY_{k}}{d\tau }& =W_{k}\ \ ,\ \ \frac{dW_{k}}{d\tau }=-\nabla \Phi
_{\varepsilon }\left( Y_{k}-X\right) \ \ ,\ \ k\in S\ .
\end{align*}

All the results included in this paper are in space dimension three and we are assuming that
the average distance between the scatterers is $d=1.$ In all the cases the
mean free path $\ell _{\varepsilon }$ is much larger than the average
distance between particles. In the case of Landau kinetic equations we
mention here some examples of potentials yielding these equations. Other
scaling limits yielding this behaviour will be discussed in \cite{NVW}. 

\begin{enumerate}
\item For potentials: $\Phi _{\varepsilon }\left( Y\right) =\Phi \left( \frac{%
\left\vert Y\right\vert }{\varepsilon }\right) ,$ with $\varepsilon
\rightarrow 0$ and $\Phi\left( r\right) $ decreasing faster than $\frac{1}{%
r^s},$ with $s >1$ we can approximate the dynamics of the
Rayleigh gas (\ref{S4E9}) using the Boltzmann-Grad time scale $\tau
=T_{\varepsilon }t$ with $T_{\varepsilon }=\frac{1}{\varepsilon ^{2}}$ using
the Boltzmann equation (cf.~\cite{BLLS}, \cite{NWL19}, \cite{S1}):%
\begin{align}
\partial _{t}f\left( v,t\right) & =Q_{B}(g,f)(v,t)  \notag \\
Q_{B}(g,f)(v)& =\int_{{\mathbb{R}}^{3}}\int_{\mathbb{S}^{2}}B\left(
(v-v_{\ast })\cdot \omega ,|(v-v_{\ast })|\right) \left( g(v_{\ast }^{\prime
})f(v^{\prime })-g(v_{\ast })f(v)\right) \;\mathrm{d}{v_{\ast }}\, \mathrm{d}{%
\omega } \ , \label{eq:opQ}
\end{align}%
where $v_{\ast }{}^{\prime },\ v^{\prime }$ are now given by the standard
formulas for the collisions, namely 
\begin{equation}
v^{\prime }=v-\omega \cdot (v-v_{\ast })\,\omega ,\qquad v_{\ast }^{\prime
}=v_{\ast }+\omega \cdot (v-v_{\ast })\,\omega .  \label{scattering}
\end{equation}

\item We can consider potentials of the form $\Phi _{\varepsilon }\left(
Y\right) =\varepsilon \Phi\left( \frac{\left\vert Y\right\vert }{L_{\varepsilon
}}\right) ,$ with $\Phi$ smooth, $1\ll L_{\varepsilon }\ll \frac{1}{\varepsilon 
}$ as $\varepsilon \rightarrow 0$ and with $\Phi\left( r\right) $ decreasing
faster as $\frac{1}{r^{s}}$ or potentials with the form $\Phi _{\varepsilon
}\left( Y\right) =\Phi\left( \frac{\left\vert Y\right\vert }{\varepsilon }%
\right) $ with $\Phi$ smooth behaving as $\Phi\left( r\right) \sim \frac{1}{r}$ as 
$r\rightarrow \infty .$ The kinetic regime is obtained in the time scale $t=%
\frac{\tau }{T_{\varepsilon }}$ with $T_{\varepsilon }=\frac{1}{\varepsilon
^{2}L_{\varepsilon }^{2}}$ in the first case and $T_{\varepsilon }=\frac{1}{%
\varepsilon ^{2}\log \left( \frac{1}{\varepsilon }\right) }$ in the second
case. In this case the resulting kinetic equation, up to some trivial
rescaling of constants in the time variable, is the Landau equation for a
Rayleigh Gas (see for instance \cite{S1}) which reads as
\begin{align}
\partial _{t}f\left( v,t\right) &= Q_{{L}}(g,f)\left( v,t\right)  \notag \\
Q_{{L}}(g,f)\left( v\right)&=\sum_{j,k}\int_{\mathbb{R}^{3}}dv_{\ast
}g\left( v_{\ast }\right) \partial _{j}\left(\kappa \left( \delta _{j,k}-%
\frac{\left( v-v_{\ast }\right) _{j}\left( v-v_{\ast }\right) _{k}}{%
\left\vert v-v_{\ast }\right\vert ^{2}}\right) \big(v_{\ast ,k}+\partial _{k}%
\big)f\left( v,t\right) \right) \ , \label{LandRayEq}
\end{align}
with $\kappa$ given as in \eqref{def:kappa}. We remark that the main
difference between the kinetic equations (\ref{LandLorMov}) and (\ref%
{LandRayEq}) is the presence of the friction term $v_{\ast ,k}f\left(
v,t\right) .$ This term yields a change in the energy of the tagged
particle, which is due to the fact that the tagged particle can exchange
energy with the scatterers, differently from the case of the Lorentz gases.
\end{enumerate}

\bigskip

\subsection{Evolution of interacting particle systems.}\label{ss:KinEqI}

As indicated in the Introduction we can obtain kinetic equations which
approximate the dynamics of the system (\ref{S1E1}) in suitable scaling
limits using as an intermediate step the kinetic limit of the dynamics of
either (\ref{S4E8}) or (\ref{S4E9}) during small macroscopic times. The
derivation of the kinetic equations is made by means of the closure equation
(\ref{ClosRel}) where $K\left[ \cdot ;g\right] $ is the kinetic kernel which
corresponds to the kinetic limit associated to (\ref{S4E8}) or (\ref{S4E9}).
The corresponding kinetic equation is then (restricting ourselves to
spatially homogeneous situations) given by (\ref{KinNonLin}).

Some of the cases of interaction potentials and the approximations of the
form (\ref{S4E8}) or (\ref{S4E9}) used in the kinetic limit are the
following ones. We write also the resulting kinetic equations which describe
the evolution of the particle system (\ref{S1E1}) as well as the asymptotic
limit in which this approximation is valid. In the case of Landau and
Balescu-Lenard equations, the specific scaling limits yielding these
kinetic equations will be made precise in Subsection \ref{sec:Approx2} and \cite{NVW}.

\smallskip

\begin{enumerate}
\item[] For potentials: $\Phi _{\varepsilon }\left( Y\right) =\Phi\left( \frac{%
\left\vert Y\right\vert }{\varepsilon }\right) ,$ with $\varepsilon
\rightarrow 0$ and $\Phi\left( r\right) $ decreasing faster than $\frac{1}{%
r^{s}},$ with $s>1$ we can approximate the evolution of the
distribution of velocities by means of the Boltzmann equation. 
\begin{align}
\partial _{t}f(v,t)& =Q_{B}(f,f)\left( v,t\right) \ \ ,\ \   \notag \\
Q_{B}(f,f)(v)& =\int_{\mathbb{R}^{3}}dv_{\ast }\int_{S^{2}}d\omega B\left(
n\cdot \omega ,\left\vert v-v_{\ast }\right\vert \right) \left[ f(v^{\prime
})f(v_{\ast }^{\prime })-f(v_{\ast })f(v)\right] ,\   \label{eq:boltzman}
\end{align}%
where $S^{2}$ is the unit sphere in $\mathbb{R}^{3}$ and $n=n\left(
v,v_{\ast }\right) =\frac{\left( v-v_{\ast }\right) }{\left\vert v-v_{\ast
}\right\vert }.$ Here $(v,v_{\ast })$ is a pair of velocities in incoming
collision configuration and $(v^{\prime },v_{\ast }^{\prime })$ is the
corresponding pair of outgoing velocities defined by the collision rule 
\begin{align}
v^{\prime }& =v+\left( \left( v_{\ast }-v\right) \cdot \omega \right) \omega
,  \label{Crule1} \\
v_{\ast }^{\prime }& =v_{\ast }-\left( \left( v_{\ast }-v\right) \cdot
\omega \right) \omega .  \label{Crule2}
\end{align}%
The collision kernel $B\left( n\cdot \omega ,\left\vert v-v_{\ast
}\right\vert \right) $ is proportional to the cross section for the
scattering problem associated to the collision between two particles.
\end{enumerate}
\medskip

We now consider weak, radially symmetric potentials $\Phi_\eps$ of the
following form: 
\begin{align}\label{eq:classpot}
\Phi_{{\varepsilon}}(x) = {\varepsilon} \Phi(x/L_{\varepsilon}).
\end{align}. 
The kinetic equations that will arise in this case are the following:
\begin{enumerate}
\item Landau equation. 
\begin{align}
& \partial _{t}f(v,t)=Q_{{L}}(f,f)(v,t)  \label{eq:landau} \\
& Q_{{L}}(f,f)(v)=\nabla _{v}\cdot \left( \int dv_{\ast }\;a(v-v_{\ast
})(\nabla _{v}-\nabla _{v_{\ast }})f(v)f_{\ast }(v)\right) ,  \notag
\end{align}%
where $a=a(V)=a(v-v_{\ast })$ denotes the matrix with components
\begin{equation}
a_{i,j}(V)=\frac{\pi ^{2}}{4}\int_{{\mathbb{R}}^{3}}dk\ k_{i}\ k_{j}\delta
(k\cdot V)|\hat{\Phi}(k)|^{2}=\frac{A}{|V|}\left( \delta _{i,j}-\hat{V}_{i}%
\hat{V}_{j}\right), \quad \;A>0  \label{eq:diffmatrixL}
\end{equation}%
which is determined by the pair interaction potential $\Phi $.

\item Balescu-Lenard equation. 
\begin{align}
& \partial _{t}f(v,t)=Q_{BL}(f,f)(v,t)  \label{eq:BalescuLenard} \\
& Q_{BL}(f,f)(v)=\nabla _{v}\cdot \left( \int dv_{\ast }\;a(v-v_{\ast
},v)(\nabla _{v}-\nabla _{v_{\ast }})f(v)f_{\ast }(v)\right) ,  \notag
\end{align}
where the matrix $a$ is given by 
\begin{equation}
a_{i,j}(V,v)=\frac{\pi ^{2}}{4}\int_{{\mathbb{R}}^{3}}dk\ k_{i}\ k_{j}\delta
(k\cdot V)\frac{|\hat{\Phi}(k)|^{2}}{|\varepsilon (k,k\cdot v)|^{2}} \ .
\label{eq:diffmatrixBL}
\end{equation}%
Here $\varepsilon $ is the so-called dielectric function, namely
\begin{eqnarray}
\varepsilon \left( k,u\right)  &=&1-(2\pi)^{\frac{3}{2}}\hat{\Phi}\left( k\right) \int_{\mathbb{R%
}^{3}}\frac{k\cdot \nabla f\left( v_{\ast },t\right) }{u+k\cdot v_{\ast }-i0}%
dv_{\ast}  \label{eq:dielectric}
\end{eqnarray}
where $\hat{\Phi}$ is the Fourier transform of the interaction potential. 
\end{enumerate}

We summarize below the kinetic limits for interacting
particle systems with potentials given as in \eqref{eq:classpot}. As before, we assume the initial configurations of
particles to be random. Particle velocities are assumed to be of order one,
and we choose the length scale so that  the number of particles per unit
volume is of order one. 
Then we obtain the following table for the kinetic equations, the
associated scaling limits and the kinetic timescale $T_{\varepsilon}$, 
depending on the choice of potential:

\begin{center}
$%
\begin{array}{|c|c|c|}
\hline
\text{{}} & 
\begin{array}{c}
\Phi (x)\in \mathcal{S}({\mathbb{R}}^{3})%
\end{array}%
\text{, or }\Phi \sim \frac{1}{|x|^{s}},|x|\geq 1,s>1 & 
\begin{array}{c}
\Phi \sim \frac{1}{|x|},\text{$|x|\geq 1$}%
\end{array}
\\ \hline
&  &  \\ 
L_{\varepsilon }=1 & \text{Landau, $T_{\varepsilon }={\varepsilon }^{-2}$} & 
\text{Landau, $T_{\varepsilon }={\varepsilon }^{-2}|\log {\varepsilon }|$}
\\ \hline
&  &  \\ 
L_{\varepsilon }={\varepsilon }^{-\alpha },\alpha \in (0,\frac{1}{3}) & 
\text{Landau }T_{\varepsilon }={\varepsilon }^{4\alpha -2} & \text{Landau, $%
T_{\varepsilon }={\varepsilon }^{4\alpha -2}|\log {\varepsilon }|$} \\ \hline
\begin{array}{c}
L_{\varepsilon }={\varepsilon }^{-\frac{1}{3}}%
\end{array}
& \text{Balescu-Lenard }T_{\varepsilon }={\varepsilon }^{-\frac{2}{3}}\text{%
{}} & \text{Landau, $T_{\varepsilon }={\varepsilon }^{-\frac{2}{3}}|\log {%
\varepsilon }|$} \\ \hline
\end{array}%
$
\end{center}

\bigskip

The characteristic time scales $T_{\ep}$ described in Sections \ref{ss:KinEqL}, \ref{ss:KinEqR}, \ref{ss:KinEqI} will be obtained in detail in \cite{NVW}.

\bigskip

\section{Strategy: Approximation by the dynamics of a tagged particle in a random force field with friction}\label{sec:ManyPartFrict}

We will now sketch the argument that we use in the companion paper \cite{NVW} to obtain the limit kinetic equation. We will discuss here only the case in which the range of the interaction is much larger than the particle distance. 
More precisely, we will consider the case in which the interaction potential 
$\Phi _{\varepsilon }\left( x\right) $ is 
\begin{equation}
\Phi _{\varepsilon }\left( x\right) =\varepsilon \Phi \left( \frac{%
\left\vert x\right\vert }{L_{\varepsilon }}\right) \ \ ,\ \ x\in \mathbb{R}%
^{3},\ \varepsilon >0  \label{S4E6}
\end{equation}%
where $\Phi \in C^{2}\left( \mathbb{R}^{3}\right)$ and the
characteristic length $L_{\varepsilon }$ is 
\begin{equation}
L_{\varepsilon }\gg 1 . \label{S4E7}
\end{equation}
We will assume that $\Phi \left( r\right) $ decreases fast enough
as $r\rightarrow \infty ,$ say exponentially, although in space dimension
three, a decay like $\frac{1}{r^{s}}$ with $s>1$ would be enough. 

\subsection{Approximation of the dynamics of a tagged particle in a Rayleigh gas}\label{RaylCompSuppPot}

Suppose that the position and velocity of a tagged particle $\left(
X,V\right) $ evolves according to (\ref{S4E9}) where the interaction
potential $\Phi _{\varepsilon }\left( x\right) $ is given as in \eqref{S4E6}, \eqref{S4E7}. 

As a first step we approximate the distribution of scatterers as the sum of a constant
density plus some gaussian fluctuations in some suitable topology. To this
end we introduce a new variable $y_{k}$ which
will be useful to describe the system on a scale where this Gaussian
approximation is valid, but that is smaller than the mean free path. In
order to keep the particle velocities of order one we introduce a new time
scale $\tilde{t}$.  
More precisely, we perform the following change of variables:
\begin{equation}
y_{k}=\frac{Y_{k}}{L_{\varepsilon}}\ \ ,\ \ \xi=\frac{X}{%
L_{\varepsilon}}\ \ ,\ \ \tilde {t}=\frac{\tau}{L_{\varepsilon}} \ .
\label{S8E6a}
\end{equation} 
Then (\ref{S4E9}) becomes:%
\begin{equation}  \label{S6E1}
\begin{aligned} \frac{d\xi }{d\tilde{t}}& =V\ \ ,\ \
\frac{dV}{d\tilde{t}}=-\varepsilon \sum_{j\in S}\nabla _{\xi }\Phi \left( \xi -y_{j}\right) \\ \frac{dy_{k}}{d\tilde{t}}& =W_{k}\ \ ,\ \
\frac{dW_{k}}{d\tilde{t}}=-\varepsilon \nabla _{y}\Phi \left( y_{k}-\xi \right) \ \ ,\ \ k\in S \ . \end{aligned}
\end{equation}

The goal is to approximate the dynamics of the scatterers by means of a
continuous density. To this end we introduce the following particle density
in the phase space:%
\begin{equation}  \label{eq:feps}
f_{\varepsilon }\left( y,w,\tilde{t}\right) =\frac{1}{\left( L_{\varepsilon
	}\right) ^{3}}\sum_{k}\delta \left( y-y_{k}\right) \delta \left(
w-W_{k}\right).
\end{equation}

We can then rewrite the first two equations of (\ref{S6E1}) as:%
\begin{equation}
\frac{d\xi }{d\tilde{t}}=V\ \ ,\ \ \frac{dV}{d\tilde{t}}=-\varepsilon \left(
L_{\varepsilon }\right) ^{3}\int_{\mathbb{R}^{3}}\nabla _{\xi }\Phi \left(
\xi -\eta \right) \rho _{\varepsilon }\left( \eta ,\tilde{t}\right) d\eta
\label{S6E2a}
\end{equation}%
where $\rho _{\varepsilon }(y,\tilde{t})=\rho \lbrack f_{\varepsilon }(\cdot
,\tilde{t})](y)$ is the spatial density introduced in \eqref{eq:spatdensity}%
. On the other hand, the second set of equations of (\ref{S6E1}) implies
that:%
\begin{equation}
\partial _{\tilde{t}}f_{\varepsilon }\left( y,w,\tilde{t}\right) +w\cdot
\nabla _{y}f_{\varepsilon }\left( y,w,\tilde{t}\right) -\varepsilon \nabla
_{y}\Phi \left( y-\xi \right) \cdot \nabla _{w}f_{\varepsilon }\left( y,w,%
\tilde{t}\right) =0 \ .  \label{S6E3}
\end{equation}

We have then reformulated (\ref{S6E1}) as (\ref{S6E2a}), (\ref{S6E3}).

We can now take formally the limit $\varepsilon \rightarrow 0.$ To this end,
notice that $f_{\varepsilon }\left( y,w,0\right) $ is of order one and it
converges in the weak topology to $g\left( w\right) .$ In order to obtain
the evolution for different rescalings of $L_{\varepsilon }$ with $\varepsilon $ we compute the asymptotic behaviour in law of $f_{\varepsilon
}\left( y,w,0\right) $ as $\varepsilon \rightarrow 0.$ We will repeatedly use the following
Gaussian approximation (\ref{S6E4}) for empirical densities associated to particle
distributions given by the Poisson measure:
\begin{equation}
\mathbb{E}\left[ \left( f_{\varepsilon }\left( y_{a},w_{a},0\right) -g\left(
w_{a}\right) \right) \left( f_{\varepsilon }\left( y_{b},w_{b},0\right)
-g\left( w_{b}\right) \right) \right] =\frac{g\left( w_{a}\right) }{\left(
	L_{\varepsilon }\right) ^{3}}\delta \left( y_{a}-y_{b}\right) \delta \left(
w_{a}-w_{b}\right) .  \label{S6E4}
\end{equation}

Assuming (\ref{S6E4}), it is natural to look for solutions of (\ref{S6E3})
with the form:%
\begin{equation}
f_{\varepsilon }\left( y,w,\tilde{t}\right) =g\left( w\right) +\frac{1}{%
	\left( L_{\varepsilon }\right) ^{\frac{3}{2}}}\zeta _{\varepsilon }\left(
y,w,\tilde{t}\right).  \label{S6E4a}
\end{equation}

Then, using the fact that the contribution to the integral $\int_{\mathbb{R}%
	^{3}}\nabla _{\xi }\Phi \left( \xi -\eta \right) \rho _{\varepsilon }\left(
\eta ,\tilde{t}\right) d\eta $ due to the term $g\left( w\right) $ vanishes,
we obtain that $\zeta _{\varepsilon }\left( y,w,\tilde{t}\right) $ solves
the following problem:

\begin{align}
\frac{d\xi }{d\tilde{t}}&=V\ \ ,\ \ \frac{dV}{d\tilde{t}}=-\varepsilon
\left( L_{\varepsilon }\right) ^{\frac{3}{2}}\int_{\mathbb{R}^{3}}\nabla
_{\xi }\Phi \left( \xi -\eta \right) \tilde{\rho}_{\varepsilon }\left( \eta ,%
\tilde{t}\right) d\eta,  \label{S6E5} \\
\partial _{\tilde{t}}\zeta _{\varepsilon }\left( y,w,\tilde{t}\right)
+&w\cdot \nabla _{y}\zeta _{\varepsilon }\left( y,w,\tilde{t}\right)
-\varepsilon \left( L_{\varepsilon }\right) ^{\frac{3}{2}}\nabla _{y}\Phi
\left( y-\xi \right) \cdot \nabla _{w}\left( g\left( w\right) +\frac{\zeta
	_{\varepsilon }\left( y,w,\tilde{t}\right)}{\left( L_{\varepsilon }\right) ^{%
		\frac{3}{2}}} \right) =0 ,  \label{S6E6}
\end{align}
where $\tilde{\rho}_{\varepsilon }\left( y,\tilde{t}\right) = \rho[%
\zeta_{\varepsilon}(\cdot,\tilde{t})](y)$ is the associated spatial density
(cf.~\eqref{eq:spatdensity}).


Then, using the fact that $L_{\varepsilon }\rightarrow \infty $ we obtain
the following limit problem formally. We write 
\begin{equation}
\theta _{\varepsilon }=\varepsilon \left( L_{\varepsilon }\right) ^{\frac{3}{%
		2}}.  \label{thetaRay}
\end{equation}

Given that the range of the interaction potentials is of order $\left\vert
y\right\vert \approx 1$ we need to have $\theta _{\varepsilon }\rightarrow 0$
as $\varepsilon \rightarrow 0$ in order to obtain a kinetic limit. On the
other hand, since $L_{\varepsilon }\rightarrow \infty ,$ making the Gaussian
approximation $\zeta _{\varepsilon }\rightharpoonup \zeta ,$ we can
approximate the problem (\ref{S6E5}), (\ref{S6E6}) as: 
\begin{equation}
\frac{d\xi }{d\tilde{t}}=V\ \ ,\ \ \frac{dV}{d\tilde{t}}=-\theta
_{\varepsilon }\int_{\mathbb{R}^{3}}\nabla _{\xi }\Phi \left( \xi -\eta
\right) \tilde{\rho}\left( \eta ,\tilde{t}\right) d\eta  \label{S6E7}
\end{equation}%
where $\tilde{\rho}\left( y,\tilde{t}\right) =\rho \lbrack \zeta (\cdot ,%
\tilde{t})](y)$ and 
\begin{align}
(\partial _{\tilde{t}}+w\cdot \nabla) _{y}\zeta \left( y,w,\tilde{t}\right)
-\theta _{\varepsilon }\nabla _{y}\Phi \left( y-\xi \right) \cdot \nabla
_{w}g\left( w\right) & =0,\quad \zeta \left( y,w,0\right) =N\left(
y,w\right) .  \label{S6E8}
\end{align}%
Using (\ref{S6E4}) we obtain:%
\begin{equation}
\mathbb{E}\left[ N\left( y,w\right) \right] =0\ \ ,\ \ \mathbb{E}\left[
\left( N\left( y_{a},w_{a}\right) \right) N\left( y_{b},w_{b}\right) \right]
=g\left( w_{a}\right) \delta \left( y_{a}-y_{b}\right) \delta \left(
w_{a}-w_{b}\right).  \label{S7E1}
\end{equation}

To prove rigorously that the term $\frac{1}{\left( L_{\varepsilon }\right) ^{%
		\frac{3}{2}}}\zeta _{\varepsilon }$ can be neglected in (\ref{S6E6}) is a
challenging mathematical problem, that will not be considered in this paper.

Due to the linearity of (\ref{S6E8})-(\ref{S7E1}) we can write $\zeta\left(
y,w,\tilde{t}\right) $ as $\zeta_{1}\left( y,w,\tilde{t}\right) +\zeta
_{2}\left( y,w,\tilde{t}\right) ,$ where:%
\begin{equation}
\partial_{\tilde{t}}\zeta_{1}\left( y,w,\tilde{t}\right) +w\cdot\nabla
_{y}\zeta_{1}\left( y,w,\tilde{t}\right) =0\ \ ,\ \ \zeta_{1}\left(
y,w,0\right) =N\left( y,w\right)  \label{S7E2}
\end{equation}%
\begin{equation}
\partial_{\tilde{t}}\zeta_{2}\left( y,w,\tilde{t}\right) +w\cdot\nabla
_{y}\zeta_{2}\left( y,w,\tilde{t}\right)
-\theta_{\varepsilon}\nabla_{y}\Phi\left( y-\xi\right)
\cdot\nabla_{w}g\left( w\right) =0\ \ ,\ \ \zeta _{2}\left( y,w,0\right) =0.
\label{S7E3}
\end{equation}

It is possible now to approximate the functions $\zeta_1, \zeta_2$ and therefore to derive an evolution equation for the tagged particle, for small macroscopic times in which the change of $V$ is
small but order one, namely:

\begin{equation}\label{eq:newEv}
\frac{dx}{dt}=w\ ,\ \ \ \frac{dw}{dt}=\eta \left( t;w\right) -\Lambda
_{g}\left( w\right) \ .
\end{equation}
Here $ \eta$ is a Gaussian random field which can be characterized by means of
\begin{equation*}
\mathbb{E}\left[ \eta \left( t_{1};w\right) \right] =0,\quad \mathbb{E}\left[
\eta \left( t_{1};w\right) \otimes \eta \left( t_{2};w\right) \right]
=D_{g}\left( w\right) \delta \left( t_{1}-t_{2}\right) .
\end{equation*}%
where:%
\begin{equation}
D_{g}\left( w\right) =\int_{0}^{\infty }\frac{ds}{s^{3}}\int_{\mathbb{R}%
	^{3}}d\eta _{1}\int_{\mathbb{R}^{3}}d\eta _{2}\nabla \Phi \left(
\eta _{1}\right) \otimes \nabla \Phi \left( \eta _{2}\right) g\left(
w+\frac{\eta _{1}-\eta _{2}}{s}\right) \ .   \label{S8E4a}
\end{equation}
Moreover, 
\begin{align}
\Lambda _{g}\left( w\right) & =-\int_{\mathbb{R}^{3}}dz \int_{\mathbb{R}%
	^{3}}du\nabla \Phi \left( z \right) \left[ \nabla g\left(
u\right) \cdot \int_{-\infty }^{0}\nabla \Phi \left( z +\left(
u-w\right) s\right) ds\right] .  \label{S8E1}
\end{align}


Therefore, if we denote the probability density describing the distribution
of the tagged particle as $f\left( x,w,t\right) $ we obtain the following
evolution equation for it%
\begin{equation}
\partial _{t}f+w\cdot \partial _{x}f=\partial _{w}\cdot \left( \frac{1}{2}%
D_{g}\left( w\right) \partial _{w}f+\Lambda _{g}\left( w\right) f\right) .
\label{KinFastDec}
\end{equation}

\subsection{Approximation of the dynamics of a tagged particle in an interacting particle system}\label{sec:Approx2}

We begin considering the case of interaction potentials with the form (\ref%
{S4E6}), (\ref{S4E7}). We consider the dynamics of a distinguished particle $\left( X,V\right) $ in an interacting particle system given as in \eqref{S1E1}. Denoting as $\left(Y_{k},W_{k}\right) ,\ k\in S$ the position and velocity of the remaining
particles of the system, using the change of variables \eqref{S8E6a}, namely $y^\pm_{k}=\frac{Y^\pm_{k}}{L_{\varepsilon}}\ , \ \xi=\frac{X}{L_{\varepsilon}}\ ,\ \tilde {t}=\frac{\tau}{L_{\varepsilon}}$, we obtain:
\begin{align}
\frac{d\xi }{d\tilde{t}}& =V\ \ ,\ \ \frac{dV}{d\tilde{t}}=-\varepsilon
\sum_{j\in S}\nabla _{\xi }\Phi \left( \xi -y_{j}\right)  \label{S9E2} \\
\frac{dy_{k}}{d\tilde{t}}& =W_{k}\ \ ,\ \ \frac{dW_{k}}{d\tilde{t}}%
=-\varepsilon \nabla _{y}\Phi \left( y_{k}-\xi \right) -\varepsilon
\sum_{j\in S}\nabla _{y}\Phi \left( y_{k}-y_{j}\right) \ \ ,\ \ k\in S  \ . \label{S9E2b}
\end{align}

We now rewrite \eqref{S9E2} as
\begin{equation}
\frac{d\xi}{d\tilde{t}}=V\ \ ,\ \ \frac{dV}{d\tilde{t}}=-\varepsilon\left(
L_{\varepsilon}\right) ^{3}\int_{\mathbb{R}^{3}}\nabla_{\xi}\Phi\left(
\xi-\eta\right) \rho_{\varepsilon}\left( \eta,\tilde{t}\right) d\eta
\label{S9E3}
\end{equation}
with $\rho_{\varepsilon}\left( y,\tilde{t}\right) =\rho[f_{\varepsilon}(%
\cdot,\tilde{t})](y)$ (cf.~\eqref{eq:spatdensity}) where $f_{\varepsilon}$  is the initial empirical measure defined in \eqref{eq:feps}.  
The second set of equations (cf.~\eqref{S9E2b}) becomes
\begin{equation}
\frac{dy_{k}}{d\tilde{t}}=W_{k}\ \ ,\ \ \frac{dW_{k}}{d\tilde{t}}%
=-\varepsilon\nabla_{y}\Phi\left( y-\xi\right) -\varepsilon\left(
L_{\varepsilon}\right)^3(\nabla_{y}\Phi *\rho_{\varepsilon})(y_k,\tilde{t})
\ ,\ \ k\in S \ .  \label{S9E4}
\end{equation}

Using (\ref{S9E4}) we can derive, arguing as in Subsection \ref%
{RaylCompSuppPot}, the following evolution equation for the particle density 
$f_{\varepsilon}:$ 
\begin{equation}
\partial_{\tilde{t}}f_{\varepsilon}\left( y,w,\tilde{t}\right) +w\cdot
\nabla_{y}f_{\varepsilon}\left( y,w,\tilde{t}\right) - {\varepsilon} \left[
\nabla_{y}\Phi\left( y-\xi\right) +\left( L_{\varepsilon }\right) ^{3}
(\nabla_{y}\Phi * \rho_{\varepsilon})( y,\tilde{t})\right] \cdot\nabla
_{w}f_{\varepsilon}\left( y,w,\tilde{t}\right) =0 \ . \label{S9E5}
\end{equation}

Using (\ref{S6E4a}), we can rewrite (\ref{S9E5}) as:%
\begin{align*}
(\partial _{\tilde{t}}& +w\cdot \nabla _{y})\zeta _{{\varepsilon }}\left(
y,w,\tilde{t}\right) \\
-& \left[ \varepsilon \left( L_{\varepsilon }\right) ^{\frac{3}{2}}\nabla
_{y}\Phi \left( y-\xi \right) +\varepsilon \left( L_{\varepsilon }\right)
^{3}(\nabla _{y}\Phi \ast \tilde{\rho}_{\varepsilon })(y,\tilde{t})\right]
\cdot \nabla _{w}\left( g\left( w\right) +\frac{1}{\left( L_{\varepsilon
	}\right) ^{\frac{3}{2}}}\zeta _{\varepsilon }\left( y,w,\tilde{t}\right)
\right) =0
\end{align*}%
where $\tilde{\rho}\left( y,\tilde{t}\right) =\rho \lbrack \zeta (\cdot ,%
\tilde{t})](y)$.  
%
We now assume that 
$$\varepsilon \left( L_{\varepsilon }\right) ^{3}\rightarrow \sigma,\ \ \ \text{as}\  \ \ \varepsilon
\rightarrow 0, \ \ \ \sigma \geq 0.$$ 
We can then approximate $\zeta _{\varepsilon }\sim \zeta $ as $%
\varepsilon \rightarrow 0$ where $\zeta$ solves the problem (cf.~\cite{NVW}): 
\begin{equation}
(\partial _{\tilde{t}}+w\cdot \nabla _{y}) \zeta \left( y,w,\tilde{t}\right) -%
\left[ \varepsilon \left( L_{\varepsilon }\right) ^{\frac{3}{2}}\nabla
_{y}\Phi \left( y-V\tilde{t}\right) +\sigma (\nabla _{y}\Phi \ast \tilde{\rho%
})\left( y,\tilde{t}\right) d\eta \right] \cdot \nabla _{w}g\left( w\right)
=0  \label{S9E6}
\end{equation}%
\begin{equation*}
\zeta \left( y,w,0\right) =N\left( y,w\right)
\end{equation*}%
where $\tilde{\rho}\left( y,\tilde{t}\right) =\rho \lbrack \zeta (\cdot ,%
\tilde{t})](y)$ and $N$ is a Gaussian noise as in (\ref{S7E1}). We have also made use of the
approximation $\xi \approx V\tilde{t}$ for times shorter than the mean free
time, and that the tagged particle starts at the origin $\xi \left( 0\right) =0$,
without loss of generality. Following the same strategy as in Section \ref{RaylCompSuppPot} we decompose $\zeta $ as 
\begin{equation}
\zeta \left( y,w,\tilde{t}\right) =\zeta _{1}\left( y,w,\tilde{t}\right)
+\zeta _{2}\left( y,w,\tilde{t}\right) .  \label{S9E7}
\end{equation}%
Here $\zeta _{1}$ solves
\begin{align}
(\partial _{\tilde{t}}+w\cdot \nabla _{y}\zeta _{1})\left( y,w,\tilde{t}%
\right) -\sigma \nabla _{w}g\left( w\right) \cdot (\nabla _{y}\Phi \ast 
\tilde{\rho}_{1})\left( \eta ,\tilde{t}\right) d\eta & =0  \label{T1E5} \\
\zeta _{1}\left( y,w,0\right) & =N\left( y,w\right)  \label{T1E6}
\end{align}%
with $\tilde{\rho}_{1}\left( \eta ,\tilde{t}\right) =\rho \lbrack \zeta
_{1}(\cdot ,\tilde{t})](y)$ (cf.~\eqref{eq:spatdensity}) and  $\zeta _{2}$ solves 
\begin{align}
(\partial _{\tilde{t}}+w\cdot \nabla _{y})\zeta _{2}\left( y,w,\tilde{t}%
\right) -\sigma \nabla _{w}g\left( w\right) \cdot (\nabla _{y}\Phi \ast 
\tilde{\rho}_{2})(y,\tilde{t})& =\varepsilon \left( L_{\varepsilon }\right)
^{\frac{3}{2}}\nabla _{y}\Phi \left( y-V\tilde{t}\right) \cdot \nabla
_{w}g\left( w\right)  \label{T1E8} \\
\zeta _{2}\left( y,w,0\right) & =0  \label{T1E9}
\end{align}%
with $\tilde{\rho}_{2}\left( \eta ,\tilde{t}\right) =\rho \lbrack \zeta
_{2}(\cdot ,\tilde{t})](y)$. 
\smallskip

The set of equations (\ref{S9E7})-(\ref{T1E9}) yields the fluctuations of
the scatterer density in the phase space. The contribution $\zeta _{1}$
contains the ``noisy" part of the fluctuations. The contribution $\zeta
_{2} $ yields the perturbation to the scatterers density induced by the
presence of the distinguished particle $\left( X,V\right) .$ It is crucial
to notice that if $\sigma $ is of order one, the resulting densities $\zeta
_{1},$ $\zeta _{2}$ would be different from those obtained for Rayleigh
gases (cf.~(\ref{S7E2}), (\ref{S7E3})). The terms proportional to $\sigma $
give the contribution due to the interactions of the scatterers with
themselves. The problems (\ref{T1E5})-(\ref{T1E6}) and (\ref{T1E8})-(\ref%
{T1E9}) are linear and can be solved using Fourier and Laplace transforms
(cf.~\cite{La, LL2}).

\begin{remark}
	\textbf{(On the Balescu-Lenard rescaling)}\label{RemDebyeRange} We have seen
	that for $\sigma $ of order one, the equation which describes the evolution
	of the density fluctuations (\ref{S9E6}) contains a term $\sigma \left(
	\int_{\mathbb{R}^{3}}\nabla _{y}\Phi \left( y-\eta \right) \tilde{\rho}%
	\left( \eta ,\tilde{t}\right) d\eta \right) \cdot \nabla _{w}g\left(
	w\right) $ yielding contributions of the same order of magnitude as the
	transport term $(w\cdot \nabla _{y}\zeta \left( y,w,\tilde{t}\right))$. This
	is the property characterizing the so called Balescu-Lenard limit. The
	possible onset of this term is the main difference between the dynamics of
	particles moving in a Rayleigh gases (where the term accounting by
	self-interactions between the scatterers does not appear) and the dynamics
	of a particle in an interacting particle system. It is important to remark
	that the scaling limit obtained above yielding the Balescu-Lenard limit is
	only valid for interaction potentials having a large but finite range. The case of Coulombian potentials will be discussed in \cite{NVW}. In the latter case a characteristic length, termed as Debye screening
	length, naturally  arises. At this length the free transport of
	particles becomes of the same order of magnitude as the self-interactions
	term.  
\end{remark}

\begin{remark}
	It is interesting to remark that in the case of solutions of the Vlasov
	equation with integrable initial data there are rigorous results in \cite{BH}
	which show that the density fluctuations can be approximated by the
	linearized Vlasov equation. Unfortunately those results cannot be applied
	for the problem considered above since the particles are distributed
	homogeneously in the whole space.
\end{remark}

\begin{remark}
	The fact that the noise term in the Balescu-Lenard equation can be obtained
	by means of the evolution of the white noise by means of a linearized
	Vlasov-Equation has been formulated in the physical literature with
	different degrees of generality (cf.~\cite{LL2, Pi87, Sc}). The most general
	formulation of this idea, so far, is the one in \cite{LL2} 
	where an arbitrary
	background of particle velocities is considered. The analysis in \cite{Pi87}
	is restricted to perturbations near the Maxwellian distribution.\ There is a
	physically very clear interpretation of Balescu-Lenard in terms of force
	acting on a cloud of particles generated by a tagged particle moving in a
	background of scatterers in \cite{Ros1, Ros2, Sc}.
\end{remark}

\smallskip

We further introduce a time scale $t=\frac{\tilde{t}}{\varepsilon
^{2}\left( L_{\varepsilon }\right) ^{3}}=\frac{\tau }{\varepsilon ^{2}\left(
L_{\varepsilon }\right) ^{4}},$ and a macroscopic length scale $x=\frac{\xi 
}{\varepsilon ^{2}\left( L_{\varepsilon }\right) ^{3}}=\frac{X}{\varepsilon
^{2}\left( L_{\varepsilon }\right) ^{4}}$. We
assume that $\varepsilon \left( L_{\varepsilon }\right) ^{\frac{3}{2}%
}\rightarrow 0$ as $\varepsilon \rightarrow 0$.  Then \eqref{S9E3} can be
approximated by a stochastic differential equation:%
\begin{equation*}
\frac{dx}{dt}=V\ \ ,\ \ \frac{dV}{dt}=\eta \left( t;V\right) +H_{g}\left(
V\right) \ .
\end{equation*}
Assuming the Penrose stability condition for plasmas, it has been shown in \cite{NVW} that the function $\eta \left( t;V\right)$ is such that
\begin{equation*}
\mathbb{E}\left[ \eta \left( t_{1};V\right) \otimes \eta \left(
t_{2};V\right) \right] =\delta \left( t_{1}-t_{2}\right) D_{g}\left( V\right) 
\end{equation*}
with
\begin{align}
D_{g}\left( V\right)  &=
{\pi\int_{\R^3} dw\int_{{\mathbb{R}}^{3}}dk\ \left[ k\otimes k\right] g(w)  \delta(k\cdot (V-w))\frac{|\hat{\Phi}(k)|^{2}}{|\Delta_{\sigma} (k,i k\cdot V)|^{2}} } \ . \label{LangBLa}
\end{align}
Moreover, the function $H_{g}\left( V\right)$ is the friction coefficient associated to Balescu-Lenard given by 
\begin{align}
H_{g}\left( V\right) =&
{\pi}\int_{\R^3} dw\int_{{\mathbb{R}}^{3}}dk\ \big(k\cdot \nabla g(w)\big) k \delta(k\cdot (V-w))\frac{|\hat{\Phi}(k)|^{2}}{|\Delta_{\sigma} (k,i k\cdot V)|^{2}}  \ .
\label{FricBLFastDec}
\end{align}
Here $\hat{\Phi}$ is the Fourier transform of the potential and
\begin{align}  \label{Psidef}
&\Psi \left( \zeta ;\theta \right) :=\int_{\mathbb{R}^{3}}\frac{\left( i \theta
\cdot \nabla _{w}g\left( w\right) \right) }{\zeta +i\left( \theta \cdot
w\right) }dw ,\\&
\Delta_{\sigma}\left( k,z\right) =1-\sigma (2\pi)^{\frac{3}{2}} \hat{\Phi}\left( k\right) \Psi
\left( \frac{z}{\left\vert k\right\vert };\frac{k}{\left\vert k\right\vert }%
\right) \ .\label{DeltaSdef}
\end{align}
{We recall that $g$ is called Penrose stable (cf.~\cite{Pe}) if the following condition is satisfied: 
\begin{equation}
\Delta _{\sigma }\left( k,z\right) \neq 0\text{ for }\func{Re}\left(
z\right) \geq 0, \ \ k\in \mathbb{R}^{3}.  \label{PenrStab}
\end{equation}
This condition guarantees that small perturbations of the stationary solutions $g(v)$ driven by the linearized Vlasov equation do not increase exponentially in time.}

We then obtain the equation yielding the evolution of the distribution
of particles $f\left( x,v,t\right) $ which is given by:%
\begin{equation}
\partial _{t}f\left( x,v,t\right) +v\cdot \partial _{x}f\left( x,v,t\right)
=\partial _{v}\left( \frac{D_{f}\left( v\right) }{2}\partial
_{v}f-H_{f}\left( v\right) f\right) \left( x,v,t\right) \ .
\label{KinBLFastDecay}
\end{equation}%
Notice that  the diffusion matrix in the velocity space
is chosen assuming that the distribution of velocities at each point $x$ is
given by $f\left( x,\cdot ,t\right) .$


\begin{remark} 
	One of the earliest derivation of the Balescu-Lenard equation \eqref{eq:BalescuLenard} was obtained by Lenard (cf.~\cite{Le}) using 	the approach proposed by Bogoliubov in \cite{Bo}. The approach introduced there is based on the BBGKY hierarchy,
	arguing that the truncated correlation function $g_2$ and truncated
	correlations of higher order stabilize on a much shorter timescale than the
	one-particle distribution function $f_1$. In our approach this corresponds
	to the stabilization of the function $H_g$, as well as of the noise $B_g$ to
	a stationary process on the short time scale introduced by Bogoliubov. The Bogoliubov approach will be discussed in Section \ref{correlations}.
\end{remark}

\subsection{The case of very slowly decreasing potentials.\label{SlowDecPot}}

In the case of Lorentz gases, it has been seen in \cite{NSV} that for
interaction potentials with the form $V_{\varepsilon }\left( x\right)
=\varepsilon V\left( x\right) $ with $V\left( x\right) \sim \frac{1}{%
\left\vert x\right\vert ^{s}}$ as $\left\vert x\right\vert \rightarrow
\infty $ with $\frac{1}{2}<s<1$ it is not possible to derive a kinetic
equation describing the evolution of a tagged particle as $\varepsilon
\rightarrow 0.$ The reason for this is that the deflections experienced by
the tagged particle have correlations of order one in distances of the order
of the mean free path. Note that in order to obtain a well defined random
force field in which the tagged particle moves, we need to assume an
electroneutrality condition. Nevertheless, the long-range correlations of
the deflections occur in spite of this assumption.

\bigskip

In the case of interacting particle systems the situation is different due
to the presence of the Debye screening. Due to the screening the interaction between particles takes place by means of a
potential with finite range. 
Differently from the case of Coulombian potentials, the Coulombian logarithm
does not appear because not all the dyadic regions contribute in the same
way. In the case $\frac{1}{2}<s<1,$ the largest contribution to the
deflections can be expected to be due to the interaction between particles
placed at distances of the order of the Debye length. This gives a picture
rather different from the one taking place in the case of Lorentz gases, but
it is also rather different from the picture in the usual Landau equation
for Coulombian potential. Seemingly the resulting kinetic equation describing
the dynamics of the interacting particle system in this case is a
Balescu-Lenard equation in which the main interaction takes place at
distances of the order of the Debye length. We plan to address this problem
in the future.

\section{A different approach to derive kinetic equations: Bogoliubov's hierarchical approach. \label{correlations}}.

We have seen in Subsection \ref{sec:Approx2} that for interacting
particle systems of the form (\ref{S1E1}) we can derive a kinetic
approximation in the case of weak interaction potentials (for the detailed computations we refer to \cite{NVW}). This was achieved by replacing the
particle distribution by a Gaussian random field evolving by
means of a Vlasov equation, and a friction term. Then the dynamics of a tagged particle in this random background approximates the evolution of a particle in the interacting particle system \eqref{S1E1} for short but macroscopic times.

The resulting kinetic
equations  are the Landau or the Balescu-Lenard
equation. There is an alternative way of deriving these equations which was
introduced by Bogoliubov (cf.~\cite{Bo}). This approach consists in deriving the
 evolution of the probability measure yielding the distribution of particles at any positive time. The crucial information contained in the Bogoliubov approach 
 is the characterization of this probability measure by means of its correlation structure. 

\smallskip

Suppose that we have a system of particles that solves (\ref{S1E1}). We
assume also that the initial distribution of particles in the phase space is
given by a generalized Poisson distribution $\nu _{g}$ 
characterized by a distribution of velocities $g$.  More precisely
\begin{equation}
\nu_{g}\left( \bigcap_{j=1}^{J}U_{B_{j},n_{j}}\right) =\prod_{j=1}^{J}\left[ 
\frac{\left\vert \int_{B_{j}}dxg\left( dv\right) \right\vert ^{n_{j}}}{%
\left( n_{j}\right) !}e^{-\int_{B_{j}}dxg\left( dv\right) }\right] 
\label{probMeas}
\end{equation}
where $\{B_{j}\}_j$ is a family of disjoint Borel set of $\mathbb{R}^{3}\times\mathbb{R}^{3}$, $n_{j}\in\mathbb{N}_{\ast}$ for each $j\in\left\{ 1,2,...,J\right\}$ and $U_{B_{j},n_{j}} $ is the event of finding $n_j$ particles in $B_j$.  We refer to \cite{NVW} for a detailed discussion. 

Our goal is to define a
family of evolutions of the probability measure $\tau\rightarrow \mathbb{P}%
_{\tau}^{\varepsilon },\ t\geq 0,$ with $\mathbb{P}_{0}^{\varepsilon }=\nu
_{g}$. The measure $\mathbb{P}_{\tau}$ is defined as follows. Suppose
that we denote as $\omega \in \Omega $ the random empirical measure associated to the initial particle configuration $(X_i,V_i)_{i\in I} $ in the phase space, i.e.
\begin{align} \label{eq:empirical}
\omega = \sum_{i\in I} \delta(X-X_i) \delta(V-V_i).
\end{align}
We denote as $U_{\tau}^{\varepsilon }$ the
evolution operator given by the evolution of the particle system (\ref{S1E1}). We will assume that $U_{\tau}^{\varepsilon }$ is well defined (and measurable) for each $\tau>0$ and we define the measure $%
\mathbb{P}_{\tau}^{\varepsilon }$ as:%
\begin{equation}
\mathbb{P}_{\tau}^{\varepsilon }=\mathbb{P}_{0}^{\varepsilon }\circ \left(
U_{\tau}^{\varepsilon }\right) ^{-1}=\nu _{g}\circ \left( U_{\tau}^{\varepsilon
}\right) ^{-1}. \label{T6E1}
\end{equation}

Due to the particle interaction, the measures $\mathbb{P}_{\tau}^{\varepsilon }$ are in general no Poisson measures. The usual way of characterizing these measures is by means
of the many-particle correlation functions $g_n$ which are obtained solving the
so-called BBGKY hierarchy:
\begin{equation}\label{BBGKY2}
\begin{aligned} 
& \partial_ \tau g_n(\tau, \alpha_n)		+\sum_{k=1}^n p_k \nabla_{q_k}g_n(\tau,\alpha_n) 		 
- \sum_{k=1}^n\int_{\R^6} \ud \eta_{n+1}\nabla_{q_k}\phi_\eps ( q_k-q_{n+1})
\nabla_{p_k}g_{n+1}(\tau,\alpha_n,\eta_{n+1})   \\
& = \sum_{k=1}^n\sum_{\ell=1}^n \nabla_{q_k}\phi_\eps (q_k-q_\ell) \nabla_{p_k}g_n(\tau,\alpha_n).  
\end{aligned}
\end{equation}

Here $\alpha_n=(\eta_1,\dots, \eta_n)$  with $\eta_i=(q_i,p_i)$ denoting the microscopic position and velocity.

Assuming that the BBGKY hierarchy allows us to characterize uniquely the
probability measures $\mathbb{P}_{t}^{\varepsilon }$ we would have:%
\begin{equation}
\mathbb{P}_{\tau}^{\varepsilon }=\mathbb{K}_{\varepsilon }\left[ g\left( \cdot
,t\right) ;\left\{ g_{k}\left( \cdot ,\tau\right) \right\} _{k=2}^{\infty }%
\right]  \label{T6E2}
\end{equation}%
where $\mathbb{K}$ is the operator that yields a random measure on the phase space $\R^3 \times \R^3$ for a given set of correlation functions $g_k$. The measure is uniquely determined by the correlation functions $g_k$ under relatively weak conditions, e.g. if there exists $C>0$ such that
\begin{align*}
	\|g_k(\cdot,\tau)\|_{L^\infty} \leq C^n.
\end{align*}
See for instance \cite{Len}, \cite{Rue} for more details on this so-called moment problem.

It was noticed by Bogoliubov that the time scale for the evolution of the
correlation functions for two or more particles $\left\{ g_{k}\left( \cdot
,\tau\right) \right\} _{k=2}^{\infty }$ 
is much shorter if $\varepsilon \rightarrow 0$ than the one for the
one-particle distribution $g\left( \cdot ,\tau\right) .$
Further, they are expected to stabilize to time-independent functions $G_k$ which only depend on $g$. Approximating the correlation
functions for two or more particles by the functions $G_k$  yields an approximation for the evolution of
the one-particle distribution $g$  on a much longer time scale.

Bogoliubov's approach is particularly useful in the derivation of Landau and
Balescu-Lenard equations. We will illustrate the use of this method in the case of interaction potentials with the form (\ref{S4E6}%
), (\ref{S4E7}). The analysis of this problem through an approximation by the dynamics of
a tagged particle with an effective friction term moving in a random force
field has been made in Subsection \ref{sec:Approx2}. In this
particular setting we can linearize the evolution equations yielding the
functions $\left\{ g_{k}\left( \cdot ,\tau\right) \right\} _{k=2}^{\infty }.$
It turns out that these linearized equations can be explicitly solved in
terms of the solutions of the corresponding linearized Vlasov equation (cf.~\cite{VW2}). In particular, the characteristic time scale for the evolution
of the functions $\left\{ g_{k}\left( \cdot ,\tau\right) \right\}
_{k=2}^{\infty }$ is the same as the one of the linearized Vlasov equation
(cf.~for instance \eqref{S9E6}). In the case of potentials with the form (%
\ref{S4E6}), (\ref{S4E7}) and since the particle velocities are of order one
we have that the characteristic time for the linearized Vlasov equation (and
the equations for the functions $\left\{ g_{k}\left( \cdot ,t\right)
\right\} _{k=2}^{\infty }$) is of order $L_{\varepsilon }.$ This time scale
is much shorter than the timescale $t$, in which the particles move along lengths of
the order of the mean free path. If the stability condition \eqref{PenrStab} holds, we have similar stability properties for
the linearized operator which yields the evolution of the functions $\left\{
g_{k}\left( \cdot ,\tau\right) \right\} _{k=2}^{\infty }.$ It then follows that
the functions $\left\{ g_{k}\left( \cdot ,\tau\right) \right\} _{k=2}^{\infty }$
approach to an equilibrium 
\begin{equation*}
G_{k}\left( \cdot ;g\right) \ \ ,\ \ k\geq 2,
\end{equation*}
which depends on the function $g.$ The equations
yielding the evolution of $\left\{ g_{k}\left( \cdot ,\tau\right) \right\}
_{k=2}^{\infty }$ contain the one-particle distribution $g$.
Since this evolution takes place on a longer timescale, we can approximate the dynamics of $g(\cdot,t)$ using $g_{k}\left( \cdot ,t\right) \simeq G_{k}\left( \cdot ;g\left(\cdot ,t\right) \right) $ for $k\geq 2$.
By (\ref{T6E2}), we obtain the following approximation for the
random  measure $\mathbb{P}_{\tau}^{\varepsilon }$
\begin{equation} 
\mu^{\varepsilon }\simeq \mathbb{K}_{\varepsilon }\left[ g\left(
\cdot ,\tau \right) ;\left\{ G_{k}\left( \cdot ;g\left( \cdot ,\tau\right) \right)
\right\} _{k=2}^{\infty }\right] :=\mathbb{H}_{\varepsilon }\left[ g\left(
\cdot ,\tau\right) \right]  \label{T6E3}
\end{equation}%
as $\varepsilon \rightarrow 0.$ 
\begin{remark}
	The measure $\mu$ given by the functions $G_k$ through the relation \eqref{T6E3} is a random measure on $\R^3\times\R^3$, but not necessarily related to a point process on this space. Hence, in contrast to the measure in \eqref{T6E2}, $\mu$ is in general not of the form \eqref{eq:empirical}. The problem whether such a representation holds is called ``full $K$-moment problem". For necessary and sufficient conditions, see for example \cite{Inf}.
\end{remark}

Finally, an approximation for the evolution of $g\left( \cdot ,t\right) $ follows by replacing $g_2$ by $G_2$ in \eqref{BBGKY2}.
In the regime described in Subsection \ref{sec:Approx2} this yields either the Landau or the Balescu-Lenard equation.

The previous argument, besides providing a different derivation of the
Landau and Balescu-Lenard equations (as well as a blueprint for a possible
rigorous derivation of these kinetic equations taking as starting point a
Hamiltonian mechanical system), provides also an interesting description of
the probability distributions which characterize the particle distributions
in macroscopic times (cf.~(\ref{T6E3})).

An exposition on possible applications of the BBGKY-method to various kinetic limits and their fluctuations can be found in \cite{Klim}.

\section{Miscellaneous}\label{sec:misc}

\subsection{Some relevant physical properties of the Vlasov-Poisson equation.}

In this Section we discuss a few properties of particle systems which are
due to their collective behaviour and can be described, if the interactions
between the particles are weak but have long range, using the Vlasov
equation. We will discuss two specific phenomena. The first one that we discuss here is
the phenomenon of screening. The second issue is the existence
of some oscillations of the particle density for particles interacting by
means of Coulomb potentials which damp very slowly. These oscillations are
usually termed as Langmuir waves. The two phenomena are well known in the
physical literature (cf.~for instance \cite{Jac, LL2}) but both of them are
relevant in the rigorous mathematical study of some problems related to the
Vlasov equations and for this reason we will describe them here.

Both phenomena can be related to the asymptotic behaviour of the dielectric
function $\epsilon \left( k,\omega \right)$ given as in \eqref{eq:dielectric} whose properties are discussed in \cite{NVW}. More precisely, Langmuir waves are related to the asymptotic
behaviour of $\epsilon \left( k,\omega \right) $ as $k\rightarrow 0.$ On the
other hand, screening properties act for arbitrary, dynamic situations.
However, they become particularly simple to describe in static situations,
i.e. for $\omega \rightarrow 0.$

\bigskip

\subsubsection*{Screening effects.}

A peculiar feature of the Coulombian potential is that even if the potential does not have a characteristic time scale there is an effective range of interaction due to a phenomenon known in the literature as Debye screening. The mathematical details related to this phenomenon will be discusses in \cite{NVW}, but we summarize here the main physical ideas behind it. 

It is interesting to remark that screening effects (i.e.~charge rearrangement
which tends to damp the effects of charges unbalances in a system) do not
require irreversible effects due to collisions. Actually the screening
properties can be described using just the (collisionless) Vlasov equation.
Screening is closely related to Landau damping and to the stability of a
medium discussed in \cite{NVW}. This fact is well
understood in the physical literature (cf.~\cite{LL2}). Some rigorous mathematical results are also available. In equilibrium situations, described by a Gibbs measure, it has been proved in \cite{BF} an exponential decay of correlations for distances larger than the screening length. In \cite{AW} it has been shown that the perturbation induced by a point charge in the stationary Vlasov Equation decays exponentially for distances larger than the screening length. 
We will discuss in this section the latter case, for the linearized Vlasov equation and illustrate the build-up of screening in time. 

\bigskip

Suppose that we consider a system of particles which can be approximated by
means of the Vlasov equation. 
We assume that the
interactions between the particles are Coulombian, and that the distribution
of particles is spatially homogeneous and characterized by a distribution of
velocities $f_{0}\left( v\right) .$ We set $h=\zeta ^{+}-\zeta ^{-}$ where $\zeta ^{+}$ and $\zeta ^{-}$ are the distribution of positive and negative charges respectively.  
Notice that $h$ denotes the perturbation of the particle density in the phase space. This
perturbation will be assumed to be sufficiently small to ensure that $h$ can
be described using the linearized Vlasov equation. We will assume that the
initial distribution of particles is $f_{0}\left( v\right) $ and that we
place a small charge at the origin at rest. We can then approximate the
evolution of the perturbation of the density of particles in the phase space 
$h$ by means of the following set of equations:%
\begin{eqnarray}
\partial _{t}h\left( x,v,t\right) +v\cdot \nabla _{x}h\left( x,v,t\right) &=&-{F}\cdot \nabla _{v}f_{0}\left( v\right) \notag  \\
{F}\left( x,t\right) &=&-\nabla _{x}\Phi \left( \left\vert
x\right\vert \right) -\int dy\int dvh\left( y,v,t\right) \nabla _{x}\Phi
\left( \left\vert x-y\right\vert \right)  \notag \\
h\left( x,v,0\right) &=&0   \label{T7E9}
\end{eqnarray}%
where $\Phi \left( \left\vert x\right\vert \right) =\frac{1}{4\pi \left\vert
x\right\vert }.$  We assume, without loss of generality, that $\int
f_{0}\left( v\right) dv=1.$ We will assume also that $f_{0}\left( v\right)
=f_{0}\left( \left\vert v\right\vert \right) .$ Notice that we can write $%
{F}\left( x,t\right) =-\nabla _{x}\varphi \left( x,t\right) ,$ with%
\begin{equation*}
\varphi \left( x,t\right) =\Phi \left( \left\vert x\right\vert \right) +\int
dy\int dvh\left( y,v,t\right) \Phi \left( \left\vert x-y\right\vert \right)\ .
\end{equation*}
Then%
\begin{equation}
-\Delta \varphi \left( x,t\right) =\delta \left( x\right) +\rho \left(
x,t\right) \ \ ,\ \ \rho \left( x,t\right) =\int h\left( x,v,t\right) dv \ .
\label{V1}
\end{equation}

Applying Duhamel's formula to (\ref{T7E9}) we obtain%
\begin{equation*}
h\left( x,v,t\right) =\int_{0}^{t}\left[ \nabla _{x}\varphi \left( x-v\left(
t-s\right) ,s\right) \cdot \nabla _{v}f_{0}\left( v\right) \right] ds
\end{equation*}%
and%
\begin{align*}
\rho \left( x,t\right) &=\int_{0}^{t}ds\int dv\left[ \nabla _{x}\varphi
\left( x-v\left( t-s\right) ,s\right) \cdot \nabla _{v}f_{0}\left( v\right) %
\right] \\&=\int_{0}^{t}\left( t-s\right) ds\int f_{0}\left( v\right) \Delta
_{x}\varphi \left( x-v\left( t-s\right) ,s\right) dv \ .
\end{align*}%
Using the change of variables $y=v\left( t-s\right) $ we have
\begin{eqnarray*}
\rho \left( x,t\right) &=&\int_{0}^{t}\frac{ds}{\left( t-s\right) ^{2}}\int
f_{0}\left( \frac{y}{t-s}\right) \Delta _{x}\varphi \left( x-y,s\right)
dy\\
&=&\Delta _{x}\left( \int_{0}^{t}\frac{ds}{\left( t-s\right) ^{2}}\int
f_{0}\left( \frac{y}{t-s}\right) \varphi \left( x-y,s\right) dy\right) \ .
\end{eqnarray*}

Using (\ref{V1}) we obtain%
\begin{align}
-\Delta _{x}\varphi \left( x,t\right) &= \delta \left( x\right) +\Delta
_{x}\left( \int_{0}^{t}\frac{ds}{\left( t-s\right) ^{2}}\int f_{0}\left( 
\frac{y}{t-s}\right) \varphi \left( x-y,s\right) dy\right) \nonumber \\&=\delta \left(
x\right) +\Delta _{x}\left( \int_{0}^{t}\frac{ds}{s^{2}}\int f_{0}\left( 
\frac{y}{s}\right) \varphi \left( x-y,t-s\right) dy\right) \ . \label{V2}
\end{align}

The equation (\ref{V2}) describes the onset of screening effects. This has
been studied in detail in \cite{Thomas}. We just consider here the steady
states of (\ref{V2}) which describe the long time asymptotics of this
equation if $f_{0}$ satisfies the stability condition \eqref{PenrStab}. The stationary solutions of (\ref{V2}) satisfy%
\begin{equation}
-\Delta _{x}\varphi _{\infty }\left( x\right) =\delta \left( x\right)
-\Delta _{x}\left( \int_{0}^{\infty }\frac{ds}{s^{2}}\int f_{0}\left( \frac{y%
}{s}\right) \varphi _{\infty }\left( x-y\right) dy\right) \ . \label{V3}
\end{equation}

We can rewrite (\ref{V3}) computing the integral $\int_{0}^{\infty }\frac{ds%
}{s^{2}}f_{0}\left( \frac{y}{s}\right) .$ To this end we write $y$ in
spherical coordinates $y=r\omega ,\ r=\left\vert y\right\vert >0,\ \omega
\in S^{2}.$ We then have $f_{0}\left( \frac{y}{s}\right) =f_{0}\left( \frac{r%
}{s}\right) ,$ since we assumed that $f_{0}$ is invariant under rotations.
Then%
\begin{equation*}
\int_{0}^{\infty }\frac{ds}{s^{2}}f_{0}\left( \frac{y}{s}\right)
=\int_{0}^{\infty }\frac{ds}{s^{2}}f_{0}\left( \frac{r}{s}\right) =\frac{1}{r%
}\int_{0}^{\infty }f_{0}\left( \xi \right) \xi ^{2}d\xi =\frac{1}{4\pi r} \ .
\end{equation*}
Making use of this identity yields
\begin{equation*}
\int_{0}^{\infty }\frac{ds}{s^{2}}\int f_{0}\left( \frac{y}{s}\right)
\varphi _{\infty }\left( x+y\right) dy=\int \frac{\varphi _{\infty }\left(
x-y\right) dy}{4\pi \left\vert y\right\vert }=\int \frac{\varphi _{\infty
}\left( y\right) dy}{4\pi \left\vert x-y\right\vert }\ .
\end{equation*}

Therefore (\ref{V3}) implies%
\begin{equation*}
-\Delta _{x}\varphi _{\infty }\left( x\right) =\delta \left( x\right) +\int
\Delta _{x}\left( \frac{1}{4\pi \left\vert x-y\right\vert }\right) \varphi
_{\infty }\left( y\right) dy=\delta \left( x\right) -\varphi _{\infty
}\left( x\right)
\end{equation*}%
whence%
\begin{equation*}
-\Delta _{x}\varphi _{\infty }\left( x\right) +\varphi _{\infty }\left(
x\right) =\delta \left( x\right)\ .
\end{equation*}

The solution of this equation is $\varphi _{\infty }\left( x\right) =\frac{%
e^{-\left\vert x\right\vert }}{4\pi \left\vert x\right\vert }.$ This
solution exhibits the expected screening property in distances of the order
of the screening length.

\bigskip

An alternative derivation of the screening properties taking as starting
point the Vlasov equation can be found in \cite{Klim} and \cite{LL2}. The
approach in those books is based in solving (\ref{T7E9}) using Fourier and
Laplace transforms in order to compute the dielectric constant $\epsilon
\left( k,\omega \right) $ which, for $\omega \rightarrow 0$ (i.e. in the
stationary regime) is proportional $\frac{1}{\left\vert k\right\vert ^{2}+1}%
, $ which is precisely the Fourier tranform of the potential $\frac{%
e^{-\left\vert x\right\vert }}{4\pi \left\vert x\right\vert }.$ It is
relevant to emphasize the link between screening properties and Landau
damping (or stability of the Vlasov medium) which plays a crucial role in
the analysis of the long time asymptotics of the solutions to (\ref{V2}).

\bigskip

Some interesting problems suggested by the previous computations are the
following ones

\begin{itemize}
\item[(i)] To study the stationary solution and its stability for the solution of
the Vlasov equation which describes the distribution of charges around a
Dirac charge moving at constant speed.

\item[(ii)] To check if the screening properties considered above take place also
for smooth potentials behaving asymptotically as $\frac{1}{\left\vert
x\right\vert}$ as $\left\vert x\right\vert \rightarrow \infty .$ Notice that
in this case the potential generated by a set of charges cannot be described
using a PDE, but the screening properties very likely depend only on the
asymptotics of the potential and not in the details of it.
\end{itemize}

\bigskip

\subsubsection*{Langmuir waves.\label{LangWaves}}

Langmuir waves are some oscillations with wavelength much larger than the
Debye screening length which take place in plasmas (cf.~\cite{Jac}, \cite%
{JacBook}). We will use also the term Langmuir waves to refer to some
oscillations with very large wavelength for particle systems interacting by
means of Coulomb-like potentials. More precisely, we consider the kinetic limit of a system of particles interacting by means of potentials $\Phi$ which are smooth and behave for large values as $\Phi
\left( s\right) \sim \frac{1}{s}$ as $s\rightarrow \infty$. Notice that we are assuming that the characteristic scale of the function $\Phi$ is much smaller than the Debye screening length.

Langmuir waves can be described using the linearized Vlasov equation. We
will illustrate the meaning of these waves using the linearized Vlasov
equation for a one-component plasma (assuming then than in the original
problem there is a background of charge to ensure electroneutrality) or a
two component plasma with two opposite signs. We then assume that the
linearized problem around a constant distribution of particles has the
following form:%
\begin{align}
\partial _{t}h\left( x,v,t\right) +v\cdot \nabla _{x}h\left( x,v,t\right)
+F\cdot \nabla _{v}f_{0}\left( v\right) &=0  \label{T7E8} \\
F\left( x,t\right) =-\int \int dydvh &\left( y,v,t\right) \left[L^2 \nabla _{x}\Phi \left( L\left\vert x-y\right\vert \right) \right],  \notag
\end{align}%
where $L\gg 1$ is the Debye length. 
We look for solutions of
this equation with the form:%
\begin{equation*}
h\left( x,v,t\right) =e^{i\left( \omega t+k\cdot x\right) }H\left( v\right)
\ \ ,\ \ k\in \mathbb{R}^{3}\ \ ,\ \omega \in \mathbb{C}
\end{equation*}%
where $H$ solves:%
\begin{align*}
i\omega H\left( v\right) +i\left( v\cdot k\right) &H\left( v\right)
+F_{0}\cdot \nabla _{v}f_{0}\left( v\right) =0 \\
F_{0} &=-\int dy\int dvH\left( v\right) e^{-ik\cdot y}\left[L^2\nabla _{y}\Phi \left( L\left\vert y\right\vert \right) \right] \ .
\end{align*}%
Hence 
\begin{equation*}
H\left( v\right) =-\frac{F_{0}\cdot \nabla _{v}f_{0}\left( v\right) }{i\left[
\omega +\left( v\cdot k\right) \right] },
\end{equation*}%
and 
\begin{equation*}
F_{0}=\int dy\int dv\frac{\left( F_{0}\cdot \nabla _{v}f_{0}\left( v\right)
\right) }{i\left[ \omega +\left( v\cdot k\right) \right] }e^{-ik\cdot y}%
\left[ \left( L\right) ^{2}\nabla _{y}\Phi \left(
L \left\vert y\right\vert \right) \right] \ .
\end{equation*}
We can rewrite the equation above in matrix form: 
\begin{equation*}
\left( I-\int dy\int dv\frac{e^{-ik\cdot y}}{i\left[ \omega +\left( v\cdot
k\right) \right] }L^2\nabla _{y}\Phi \left(
L \left\vert y\right\vert \right) \otimes \nabla
_{v}f_{0}\left( v\right) \right) F_{0}=0
\end{equation*}%
for some $F_{0}\in \mathbb{C}^{3}.$ This yields the following eigenvalue
problem:%
\begin{equation}
\det \left( I-\int dy\int dv\frac{e^{-ik\cdot y}}{i\left[ \omega +\left(
v\cdot k\right) \right] }L^2\nabla _{y}\Phi \left(
L \left\vert y\right\vert \right) \otimes \nabla _{v}f_{0}\left( v\right) \right) =0  \label{StabEqu}
\end{equation}%
where $k\in \mathbb{R}^{3}$ is given. The stability condition \eqref{PenrStab} implies that the solutions of the equation (\ref{StabEqu}) are contained
in the half-plane $\left\{ \func{Im}\left( \omega \right) >0\right\} .$\
However, in the very relevant case of Coulombian potentials $\Phi \left(
\left\vert y\right\vert \right) =\frac{1}{\left\vert y\right\vert }$ it
turns out that there is a root $\omega =\omega \left( k\right) $ which
converges to a real value $\Omega _{0}\neq 0$ as $\left\vert k\right\vert
\rightarrow 0$. In \cite{NVW} we show that $\func{Im}\left( \omega \left( k\right) \right)
>0$ for all $k\in \mathbb{R}^{3}$. However, the following asymptotic formula
holds if we assume that the distribution of particle velocities $g$ is
Maxwellian%
\begin{equation}
\func{Im}\left( \omega \left( k\right) \right) \sim -c_{0}\omega _{p}\left( 
\frac{k_{D}}{k}\right) ^{3}\exp \left( -\frac{k_{D}^{2}}{2\left\vert
k\right\vert ^{2}}\right) \text{\ \ as }\frac{\left\vert k\right\vert }{k_{D}%
}\rightarrow 0  \label{T6E7}
\end{equation}%
where $k_{D}=\frac{\omega _{p}}{\left\langle v^{2}\right\rangle },$ $%
\left\langle v^{2}\right\rangle $ is the variance of the velocity, $\omega
_{p}=\sqrt{4\pi}$  and $c_{0}=\sqrt{\frac{\pi}{8}}$ (cf.~\cite{JacBook},
Chapter 10). Formula (\ref{T6E7}) is valid for the Coulomb potential $\Phi
\left( s\right) =\frac{1}{\left\vert s\right\vert },$ but similar formulas
hold for potentials $\Phi \left( s\right) $ behaving asymptotically as $%
\frac{1}{\left\vert s\right\vert }$ as $\left\vert s\right\vert \rightarrow
\infty .$ The main consequence of (\ref{T6E7}) is the existence of solutions
of (\ref{T7E8}) with very large wavelength (compared with the Debye length)
and damping in time extremely slowly. More precisely, this very slow damping of solutions are known in the physical literature as Langmuir waves. A consequence of this is the existence
of solutions to the linear problem (\ref{T7E8}) which converge to
equilibrium very slowly in suitable Sobolev spaces. This is a result that
has been rigorously proved in \cite{GS1}, \cite{GS2} for the one-dimensional
and radial version of (\ref{T7E8}) with Coulombian interactions. The
extremely slow damping of very large wavelengths plays a crucial role in the
stabilization of the correlation function for Coulombian potentials, as has
been discussed in \cite{VW2} both in the case of Maxwellian and
non-Maxwellian distributions of particle velocities. The asymptotics of
slowly damping, long wavelength waves arising in the Vlasov-Poisson system
has been rigorously studied also in~\cite{BoPo19}.

For general non Coulombian potentials, Langmuir waves do not exist. More
precisely, for general interaction potentials we have $\omega \left(
k\right) \rightarrow 0$ as $\left\vert k\right\vert \rightarrow 0.$

\bigskip

It is worth to mention that the analysis of the eigenvalue condition \eqref{StabEqu} is equivalent to the study of the function 
\begin{equation}
\Delta_L \left( k,z\right) =1-\frac{2\left( 2\pi \right) ^{%
\frac{3}{2}}}{ L ^{2}}\hat{\Phi}\left( \frac{k}{%
L}\right) \Psi \left( \frac{z}{\left\vert k\right\vert },%
\frac{k}{\left\vert k\right\vert }\right)  \quad \text{where}\ \ \ \Psi \left( z,k\right) =\int_{\mathbb{R}^{3}}dw_{0}\frac{
ik\cdot \nabla _{w}g\left( w_{0}\right)  }{\left( z+i\left( k\cdot
w_{0}\right) \right) } \ . \label{T8E6a}
\end{equation} 
This function plays, for Coulombian potentials, the same role of the function $\Delta_\sigma$ in \eqref{DeltaSdef}-\eqref{Psidef} in the analysis of particle systems interacting by means of potentials with the form \eqref{S4E6}-\eqref{S4E7}. 

Notice that the function $\left( \Delta _{L}\left( k,z\right)
\right) ^{-1}$ is analytic in the variable $z$ as long as $\Delta
_{L}\left( k,\cdot \right) $ does not have a zero. On the other
hand, the existence of Langmuir waves implies the existence of one zero of $%
\Delta _{L}\left( k,z\right) $ with $z=\omega i$ and $\omega $
satisfying (\ref{T6E7}). Therefore, in the case of Coulombian potentials the
function $\left( \Delta _{L}\left( k,z\right) \right) ^{-1}$ is
analytic in the $z$ variable in a region with the form $\left\{ \func{Re}%
\left( z\right) \geq -C\exp \left( -\frac{a}{\left\vert k\right\vert ^{2}}%
\right) \right\} $ for some positive constants $C$ and $a.$ Moreover, the
size of this region of analiticity is rather optimal. In particular, it is
not possible to obtain analyticity of the function $\left( \Delta
_{L}\left( k,\cdot \right) \right) ^{-1}$ in a region with the
form $\left\{ \func{Re}\left( z\right) \geq -C\left\vert k\right\vert
\right\} $ as in the case of long range, fast decaying potentials considered
in Section \ref{sec:Approx2}. The main consequence of having an
analiticity region for $\left( \Delta _{L}\left( k,z\right)
\right) ^{-1}$ is that the solutions of the Vlasov equation for Coulomb-like potentials (cf.~\eqref{T7E8}) with characteristic scale larger than the Debye screening length decay very slowly as time goes to infinity (Langmuir waves). 
In particular it would be relevant to understand what is the effect of such
long wave perturbations in the solutions of the initial value problems for
the linearized Vlasov equation yielding the analogous of the friction term \eqref{FricBLFastDec} and the noise term (\ref{LangBLa}) for Coulomb-like potentials.

\medskip

We further observe that in the case of Coulombian interaction potentials if
the initial particle distribution has correlations at distances larger than
the Debye screening length we can expect these correlations to vanish on a
very long time scale. In particular, the existence of a kinetic limit in
this case is not clear a priori. 
Another interesting issue is to determine if there are Langmuir waves for
potentials with the form \eqref{S4E6}-\eqref{S4E7}, or if Langmuir waves are restricted to the case of Coulombian potentials.

\medskip

\subsection{Reformulation of the dynamics of a particle in a
	Rayleigh-Boltzmann gas as the dynamics of a particle in a Boltzmann random
	force field with nonelastic collisions. \label{BotzInel}}

We have seen in Subsection \ref{RaylCompSuppPot} that the dynamics of a tagged
particle in a Rayleigh gas under some smallness assumptions on the
interaction potentials can be approximated by means of the dynamics of a
tagged particle in a random force field with a friction coefficient (cf.~\eqref{eq:newEv}-\eqref{S8E1}). Both the friction coefficient and the
random force field are due to the combined effect of many scatterers. Due to
this it is remarkable that, as we will see in this Subsection, a similar
decomposition of the forces acting on a tagged particle moving in a
Rayleigh-Boltzmann gas, can be made at least in the case in which the
interaction between the tagged particle and the scatterers takes place by
means of hard-sphere potentials. More precisely, we will show that the
dynamics of a tagged particle in such a situation is equivalent to the
dynamics of a particle in the random force field generated by a set of
moving scatterers, whose dynamics is not affected by the tagged particle,
but where the collisions between the tagged particle and the scatterers are
non elastic. Such non elastic collisions play a role analogous to the
friction term in \eqref{eq:newEv}.

To prove our claim we must study first the elastic collisions between a
tagged particle and one scatterer. We will assume that the tagged particle
and the scatterer have the same mass and the same radius $a.$ We will denote
the velocities of the tagged particle before and after the collision as $%
w_{1},w_{1}^{\prime }\in \mathbb{R}^{3}$ respectively and the velocities of
the scatterer before and after the collision as $w_{2},w_{2}^{\prime }\in 
\mathbb{R}^{3}.$ We denote as $x_{1},x_{2}$ the position of the tagged
particle and scatterer at the collision time respectively. Then $\left\vert
x_{1}-x_{2}\right\vert =2a.$ We write:%
\begin{equation}
\zeta =\frac{\left( x_{1}-x_{2}\right) }{\left\vert x_{1}-x_{2}\right\vert }\ .
\label{T4E9}
\end{equation}
The conservation of momentum and energy yields:%
\begin{equation*}
w_{1}=w_{1}+w_{2}=w_{1}^{\prime }+w_{2}^{\prime }\ ,\ \ \left( w_{1}\right)
^{2}+\left( w_{2}\right) ^{2}=\left( w_{1}^{\prime }\right) ^{2}+\left(
w_{2}^{\prime }\right) ^{2}\ .
\end{equation*}
Then the exchange of momentum between the particles will be assumed to be
proportional to $\zeta .$ Moreover, we can analyze the collision in the
coordinate system in which $w_{2}=0.$ Then:%
\begin{equation}
w_{1}^{\prime }-w_{1}=-w_{2}^{\prime }=\lambda \zeta  \label{T5E2}
\end{equation}%
for some $\lambda \in \mathbb{R}$. Then, the conservation of energy  implies
\begin{equation*}
\left( w_{1}\right) ^{2}+\left( w_{2}\right) ^{2}=\left( w_{1}+\lambda \zeta
\right) ^{2}+\lambda ^{2}\left\vert \zeta \right\vert ^{2}=\left(
w_{1}\right) ^{2}+2\lambda \left( w_{1}\cdot \zeta \right) +2\lambda ^{2}
\end{equation*}%
and using that $\left\vert \zeta \right\vert =1$ we obtain:%
\begin{equation}
\lambda =-\left( w_{1}\cdot \zeta \right)  \label{T5E3}
\end{equation}%
whence:%
\begin{equation}
w_{1}^{\prime }=w_{1}-\left( w_{1}\cdot \zeta \right) \zeta \ \ ,\ \
w_{2}^{\prime }=\left( w_{1}\cdot \zeta \right) \zeta \ . \label{T5E1}
\end{equation}

We now prove that we can obtain an equivalent dynamics for the tagged
particle assuming that the scatterer does not modify its trajectory in the
collision (i.e. the random force field is not affected by the tagged
particle), but including in the collision an additional term which makes the
collision inelastic.

We will denote the velocities of the tagged particle and the scatterer
before the collision as $v_{1},v_{2}$ respectively and after the collision
as $v_{1}^{\prime },v_{2}^{\prime }.$ Then $v_{2}=v_{2}^{\prime }=0$ and $%
v_{1}=w_{1},\ v_{1}^{\prime }=w_{1}^{\prime }.$ We assume that during the
collision an impulse $I=w_{1}^{\prime }-w_{1}$ is transmitted to the tagged
particle. Then, taking into account (\ref{T5E2}), (\ref{T5E3}) we obtain:%
\begin{equation*}
I=-\left( w_{1}\cdot \zeta \right) \zeta \ .
\end{equation*}
Notice that this impulse can be thought as a friction term because the
energy of the tagged particle is reduced. Indeed, we have:%
\begin{equation*}
\left( v_{1}^{\prime }\right) ^{2}-\left( v_{1}\right) ^{2}=-\left(
w_{1}\cdot \zeta \right) ^{2}\leq 0 \ .
\end{equation*}

Moreover, the collision is kinematically possible, in spite of the fact that
the velocity of the scatterer is not modified. Indeed, the collision is
kinematically possible if we have $\left( w_{1}^{\prime }\cdot \zeta \right)
\leq 0.$ Actually we have:%
\begin{equation*}
w_{1}^{\prime }\cdot \zeta =w_{1}\cdot \zeta -\left( w_{1}\cdot \zeta
\right) \left\vert \zeta \right\vert ^{2}=0
\end{equation*}%
which confirms that the collision is possible. Notice that this formula
implies in addition that in a coordinate system in which $%
w_{2}=0$ the tagged particle (i.e. particle $1$) moves perpendicularly to the
vector $\zeta $ connecting both particles immediately after the collision.

It is natural to ask if it is possible to reformulate the problem of the
collision between two particles which interact by means of a potential that
depends only on their distance. The goal would be to check if the deflection
experienced by one of the particles (the tagged particle) is the same as the
one experienced by a tagged particle which interacts with a scatterer which
affects the dynamics of the tagged particle but is not affected by it. This could be achieved by
including an additional friction term depending only on the velocity and the
distance between the particles. More precisely the problem is the following.
We have two interacting particles which will be assumed to have the same mass
that can be assumed to take the value one. The interaction potential is $\Phi\left(
\left\vert X\right\vert \right)$. Then, the elastic collision problem is: 
\begin{eqnarray*}
	\frac{dX_{1}}{dt} &=&V_{1}\ \ ,\ \ \frac{dX_{2}}{dt}=V_{2} \\
	\frac{dV_{1}}{dt} &=&-\nabla_{X_{1}}\Phi\left( \left\vert
	X_{1}-X_{2}\right\vert \right) \ \ ,\ \ \frac{dV_{2}}{dt}=-\nabla_{X_{2}}\Phi\left( \left\vert X_{2}-X_{1}\right\vert \right) \ .
\end{eqnarray*}

Suppose that we solve a collision problem for the previous system of
equations imposing $V_{1}\left( -\infty \right) =V_{1,in}$ and assuming that
the collision between two particles is characterized by an impact parameter $%
b\in \mathbb{R}^{3}$. The problem is the following one: is it possible to
find a friction force $\Lambda \left( W;\left\vert
Y\right\vert \right)$, where $W=V_{1}-V_{2}$ and $Y=X_{1}-X_{2}$, such that the solution of the following
problem%
\begin{equation*}
\frac{dY}{dt}=W\ \ ,\ \ \frac{dW}{dt}=-\nabla_{X_{1}}\Phi\left(
\left\vert Y\right\vert \right) +\Lambda \left( W;\left\vert
Y\right\vert \right) \ ,\ \ W\left( -\infty \right) =V_{1,in}
\end{equation*}%
with the same impact parameter $b$ yields $W\left( \infty \right) =V_{1,out}$ ? 
 
\smallskip

Notice that, differently from the problem considered in Subsections
\ref{RaylCompSuppPot}, \ref{sec:Approx2} where we obtain an approximation of all the
forces acting on a tagged particle by means of a friction term plus the
evolution in a random force field, the description of the collisions given
above is just a mathematical construction without any real physical
significance. Nevertheless, the possibility of this reformulation of the
collision problem has some independent interest.

\section{Conclusions.}\label{sec:concl}

In this paper we have examined the precise conditions under which we can
approximate the dynamics of many particle systems or tracer particles in
Rayleigh gases by means of kinetic equations. In particular we have focused
mostly in the case in which the particles interact by means of weak, but
long-range potentials, including potentials behaving for long distances as
the Coulomb potential. In this type of problems the resulting kinetic
equations are the classical Landau and Balescu-Lenard equations. In the case
of system with Coulombian interactions a distinct logarithmic correction
known as Coulombian logarithm appears in the formula of the mean free path
and in the formula of the macroscopic time scale.

We have examined in detail how to derive formally both types of kinetic
equations (Landau and Balescu-Lenard), approximating the dynamics of the
tracer particles or the particles of the system by means of the dynamics of
particles in random force fields. In the kinetic limit under consideration
such a dynamics can be approximated as the dynamics of a particle with a
friction coefficient which is affected by a random force field, which is not
affected by the tagged particle itself. These results allow to obtain
nonlinear evolution equations describing the evolution of many interacting
particle systems in suitable asymptotic limits.

A general idea that we have used repeatedly to obtain kinetic approximations
of the particle systems which interact by means of weak interactions is to
approximate the dynamics of each particle by means of the dynamics of a
particle interacting with a random force field. It then turns out that the
dynamics of the tagged particle can be reformulated as the dynamic of a
particle in which two forces act, namely a friction force and a random force
field which is not affected by the tagged particle itself.

\smallskip

We have also discussed the relation between the derivations of Landau and
Balescu-Lenard equations given in this paper and the traditional derivation
obtained using the methods of \cite{Ba2,Bo,Gue} and Lenard based in the BBGKY hierarchies.
We have also analysed several phenomena taking place in many
interacting particle systems due to their collective behaviour, like
screening as well as the slow decay of some long-range oscillations known as
Langmuir waves.

\smallskip

\textbf{Acknowledgment. }The authors acknowledge support through the CRC
1060 \textit{The mathematics of emergent effects }at the University of Bonn
that is funded through the German Science Foundation (DFG), as well as the
support of the Hausdorff Research Institute for Mathematics (Bonn), through the Junior
Trimester Program on Kinetic Theory. R.W. acknowledges support  of Universit\'e de Lyon through the IDEXLYON Scientific Breakthrough Project ``Particles drifting and propelling in turbulent flows", 
and the hospitality of the UMPA ENS Lyon.

\bigskip


\bigskip


\begin{thebibliography}{99}

\bibitem{AlVi} Alexandre, R., Villani, C.: On the Landau approximation in
plasma physics. \emph{Ann. Inst. Henri Poincar\'e Anal. Nonlin\'eaire} 
\textbf{21}(1), 61--95, 2004

\bibitem{AW} Arroyo-Rabasa, A., Winter, R.: Debye screening for the stationary Vlasov-Poisson Equation in interaction with a point charge. ArXiv preprint: 2005.09764, 2020


\bibitem{Ba1}{Balescu, R.: Equilibrium and Nonequilibrium Statistical
Mechanics. John Wiley \& Sons, New-York, 1975}

\bibitem{Ba2}{Balescu, R.: Statistical Mechanics of Charged Particles.
Monographs in Statistical Physics and Thermodynamics, vol. 4. Interscience
Publishers, London, 1963}

\bibitem{BNP}
Basile, G., Nota, A. and Pulvirenti, M.:
A diffusion limit for a test particle in a random distribution of scatterers.
\emph{J. Stat. Phys.},  \textbf{155}(6), 1087--1111, 2014

\bibitem{BNPP}
Basile, G., Nota, A. Pezzotti, F., Pulvirenti, M.: Derivation of the Fick's law for the Lorentz model in a low density regime. \emph{Commun. Math. Phys.} \textbf{336}(3), 1607--1636, 2015


\bibitem{BF} Brydges D. C., Federbush, P.: Debye screening. 
\emph{Comm. Math. Phys.} \textbf{73}(3), 197--246, 1980

\bibitem{BLLS} van Beijeren, H., Lanford, O.E., Lebowitz, J.L., Spohn, H.: 
\emph{J. Stat. Phys.} \textbf{22}(2), 237--257, 1980


\bibitem{BoPo19} Bobylev, A.V., Potapenko, I.F.: Long Wave Asymptotics for
the Vlasov-Poisson-Landau Kinetic Equation. \emph{J. Stat. Phys.} \textbf{%
175}, 1--18, 2019

%

\bibitem{BGS} Bodineau, T., Gallagher, I., Saint-Raymond, L.: The Brownian
motion as the limit of a deterministic system of hard-spheres. \emph{Invent.
Math.} \textbf{203}, 493--553, 2016

\bibitem{Bo} Bogolyubov, N. N.: Problems of a dynamical theory in
Statistical Physics. Moscow State Technical Press, 1946, in Russian; English
translation in Studies in Statistical Mechanics I, edited by J. de Boer and
G. E. Uhlenbeck, part A, Amsterdam: North--Holland, 1962

\bibitem{BH} Braun,W. and Hepp, K.: The Vlasov dynamics and its fluctuations
in the $1/N$ limit of interacting classical particles. \emph{Comm. Math.
Phys.} \textbf{56}(2), 101--113, 1977


\bibitem{Ce} Cercignani, C.: Theory and Application of the Boltzmann
Equation, Scottish Academic Press, Edinburgh--London, 1975



\bibitem{Ch1} Chavanis, P.H. Kinetic theory of spatially homogeneous systems with long-range interactions: I. General results. \emph{Eur. Phys. J. Plus} \textbf{127}, 19, 2012

\bibitem{Ch2} Chavanis, P. H.: Kinetic theory of spatially homogeneous systems with long-range interactions: II. Historic and basic equations. \emph{Eur. Phys. J. Plus} \textbf{128}, 126, 2013.

\bibitem{Ch3} Chavanis, P.H. Kinetic theory of spatially homogeneous systems with long-range interactions: III. Application to power law potentials, plasmas, stellar systems, and to the HMF model. \emph{Eur. Phys. J. Plus}  \textbf{128}, 128, 2013




\bibitem{DR} Desvillettes, L. and Ricci, V.: A rigorous derivation of a linear
kinetic equation of Fokker--Planck type in the limit of grazing collisions.
\emph{J. Stat. Phys.} \textbf{104}, 1173--1189, 2001

\bibitem{DGL} D\"urr, D., Goldstein, S. and Lebowitz, J.: Asymptotic motion of
a classical particle in a random potential in two dimensions: Landau model.
\emph{Comm. Math. Phys.} \textbf{113}, 209--230, 1987

\bibitem{DP}Desvillettes, L. and Pulvirenti, M.: The linear Boltzmann
equation for long--range forces: a derivation from particle systems.
\emph{Models Methods Appl. Sci. } {\textbf{9}}, 1123--1145, 1999



\bibitem{DV1} Desvillettes, L. and Villani, C.: On the spatially homogeneous
Landau equation for hard potentials. II. H-theorem and applications, \emph{%
Comm. Partial Differential Equations} \textbf{25} 261--298, 2000

\bibitem{DV2} Desvillettes, L. and Villani, C.: On the spatially homogeneous
Landau equation for hard potentials. I. Existence, uniqueness and
smoothness, \emph{Comm. Partial Differential Equations} \textbf{25},
179--259, 2000


%

\bibitem{Ein05} Einstein, A.: Die von der molekularkinetischen Theorie der W%
\"{a}rme geforderten Bewegung von in ruhen-den Fl\"{u}ssigkeiten
suspendierten Teilchen, \emph{Ann. Phys.} \textbf{17}, 549--560, 1905 

\bibitem {G} Gallavotti, G.: Grad-Boltzmann limit and Lorentz's Gas.
\emph{Statistical Mechanics. A short treatise.} Appendix 1. A2. Springer,
Berlin, 1999

\bibitem{GST} Gallagher, I., Saint Raymond, L., Texier, T.: From Newton to Boltzmann: hard spheres and short-range potentials. \emph{Z\"{u}rich Adv. Lect. in Math. } Ser. \textbf{18}, EMS, 2014

\bibitem{GS1} Glassey, R., Schaeffer, J.: Time decay for solutions to the
linearized Vlasov equation. \emph{Transp. Theory Stat. Phys.} \textbf{23}%
(4), 411--453, 1994

\bibitem{GS2} Glassey, R., Schaeffer, J.: On time decay rates in Landau
damping. \emph{Comm. Partial Different. Eqs.} \textbf{20} (3--4), 647--676,
1995

\bibitem{Gou} Goudon, T.: On Boltzmann equations and Fokker-Planck
asymptotics: influence of grazing collisions, \emph{J. Statist. Phys.} 
\textbf{89}, 751--776, 1997

\bibitem{Gue} Guernsey, R.L. : PhD thesis, University of Michigan, 1960 (unpublished)

\bibitem{Inf} Infusino, M.: The full moment problem on subsets of probabilities and point configurations. \emph{J. Math. Anal. Appl}  \textbf{483} (1), 123551, 2020


\bibitem{Jac} Jackson, J. D.: Longitudinal plasma oscillations. \emph{Plasma
Physics (J. Nucl. Energy Pt. C)} \textbf{1}(4), 171, 1960

\bibitem{JacBook} Jackson, J. D.: \emph{Classical Electrodynamics.} (2rd
Ed.). John Wiley\&Sons, New York, 1975

\bibitem {KP}Kesten, H., Papanicolaou, G.: A limit theorem for stochastic
acceleration. \emph{Comm. Math. Phys.} \textbf{78}, 19--63, 1981

\bibitem{Klim} Klimontovich, Y. L.: \emph{Kinetic Theory of Nonideal Gases
and Nonideal Plasmas}, Pergamon Press, 1982

\bibitem{Ku66} Kubo, R.: The fluctuation-dissipation theorem, \textit{%
Reports on Progress in Physics} \textbf{29}(1), 255--284, 1966 
 
\bibitem{Lan1} Lancellotti, C.: On the fluctuations about the Vlasov limit
for $N$-particle systems with mean field interactions. \emph{J. Stat. Phys.}  
\textbf{136}(4), 643--665, 2009 


\bibitem{Lan2} Lancellotti, C.: Time-asymptotic evolution of spatially
uniform Gaussian Vlasov fluctuation fields. \emph{J. Stat.Phys.} \textbf{163}%
(4), 868--886, 2016

\bibitem{La} Landau, L.D.: Die kinetische Gleichung f\"ur den Fall
Coulombscher Wechselwirkung. \emph{Phys. Zs. Sow. Union} \textbf{10}(154),
1936  


\bibitem{LL1} Landau, L.D., Lifshitz, E.M.: Mechanics. \emph{Course of
Theoretical Physics}. vol.\textbf{1}. Pergamon press, Oxford, 1960

\bibitem{LL2} Landau, L.D., Lifshitz, E.M.: Physical Kinetics. \emph{Course
of Theoretical Physics}. vol.\textbf{10}. Pergamon press, Oxford--Elmsford,
N.Y., 1981
 
\bibitem{Len} Lenard, A.: Correlation  Functions and the Uniqueness of the State in  Classical  Statistical Mechanics. \emph{Comm. Math. Phys.} \textbf{30} (1), 35-44, 1973


\bibitem{Le} Lenard, A.: On Bogoliubov's kinetic equation for a spatially
homogeneous plasma. \emph{Ann. Phys. }\textbf{10}, 390--400, 1960

\bibitem{Lo} Lorentz, H.A.: The motion of electrons in metallic bodies. 
\emph{Proc. Acad. Amst.} \textbf{7}, 438--453, 1905

\bibitem{LT} Lutsko, C., T\'oth, B.: Invariance Principle for the Random Lorentz Gas - Beyond theBoltzmann-Grad Limit. \emph{Comm. Math. Phys.}, \textbf{379}, 589--632, 2020

\bibitem{MN} Marcozzi, M., Nota, A.: Derivation of the linear Landau equation
and linear Boltzmann equation from the Lorentz model with magnetic field.
\emph{J. Stat. Phys.} \textbf{162}, 1539--1565, 2016

\bibitem{NSV} Nota, A., Simonella, S., Vel\'azquez, J.L.: On the theory of
Lorentz gases with long-range interactions. \emph{Rev. Math. Phys.} \textbf{30} (3), 1850007, 2018

\bibitem{NWL19} Nota, A., Winter, R., Lods, B.: Kinetic Description of a
Rayleigh Gas with Annihilation. \emph{J. Stat. Phys.} \textbf{176}(6),
1434--1462, 2019

\bibitem{NVW} Nota, A., Vel\'azquez, J.L., Winter, R.: Interacting particle systems
with long-range interactions:  tagged particles in random fields approximation.
2020 


\bibitem{Pi87} Piasecki, J., Szamel, G. Stochastic dynamics of a test
particle in fluids with weak long-range forces. \emph{Physica A }\textbf{123}
(1-2), 114--122, 1987

\bibitem{Pe} Penrose, O.: Electrostatic instabilities of a uniform
non-Maxwellian plasma. \emph{Phys. Fluids} \textbf{3}, 258--265, 1960

\bibitem{PSS} Pulvirenti, M., Saffirio, C., Simonella, S.:
On the validity of the Boltzmann equation for short range potentials. \emph{Rev. Math. Phys.} \textbf{26}(02), 1450001, 2014 



\bibitem{Ros1} Rostoker, N.: Superposition of Dressed Test Particles, \emph{Phys. Fluids} \textbf{7}, 479, 1964 

\bibitem{Ros2} Rostoker, N.: Fluctuations of a plasma (I), \emph{Nuclear Fusion}, \textbf{1}(2), 1961

\bibitem{Rue} Ruelle, D.: Statistical  mechanics,  rigorous  results. W. A. Benjamin Inc., New York, 1969


\bibitem{Sc} Schram, P. P. J. M.: Kinetic Theory of Gases and Plasmas.
Fundamental Theories of Physics \textbf{46}, Springer, 1991

\bibitem{S} Spohn, H.: On the Integrated Form of the BBGKY Hierarchy for
Hard Spheres, 1985, available on arXiv:math-ph/0605068

\bibitem{S1} Spohn, H.: Large scale dynamics of interacting particles. Texts
and Monographs in Physics, Springer Verlag, Heidelberg, 1991

\bibitem{S2} Spohn, H.: Kinetic equations from Hamiltonian dynamics:
Markovian limits. \emph{Rev. Mod. Phys.} \textbf{53}, 569--615, 1980

\bibitem {S3} Spohn, H.: The Lorentz flight process converges to a random flight
process. \emph{Comm. Math. Phys.} \textbf{60}, 277--290, 1978




\bibitem{VW2} Vel\'{a}zquez, J. J. L., Winter, R.: The Two-Particle
Correlation Function for Systems with Long-Range Interactions. \emph{J.
Stat. Phys.} \textbf{173} (1), 1--41, 2018

%
  
\bibitem{Thomas} Vogt, T.: Debye screening for the Vlasov-Poisson system.
Master Thesis. Rheinischen Friedrich-Wilhem Universit\"at Bonn, 2015
\end{thebibliography}
\end{document}